# Evolution of 3D Printing Methods and Materials for Electrochemical Energy Storage

Vladimir Egorov[1], Umair Gulzar[1], Yan Zhang[1], Siobhán Breen[1], and Colm O'Dwyer[1,2,3,4]*

[1]*School of Chemistry, University College Cork, Cork, T12 YN60, Ireland*

[2]*Tyndall National Institute, Lee Maltings, Cork, T12 R5CP, Ireland*

[3]*AMBER@CRANN, Trinity College Dublin, Dublin 2, Ireland*

[4]*Environmental Research Institute, University College Cork, Lee Road, Cork T23 XE10, Ireland*

**Additive manufacturing has revolutionized the building of materials direct from design, allowing high resolution rapid prototyping in complex 3D designs with many materials. 3D printing has enabled high strength damage-tolerant structures, bioprinted artificial organs and tissues, ultralight metals, medicine, education, prosthetics, architecture, consumer electronics, and as a prototyping tool for engineers and hobbyists alike. 3D printing has emerged as a useful tool for complex electrode and material assembly method for batteries and electrochemical supercapacitors in recent years. The field initially grew from extrusion based methods such as fused deposition modelling, and quickly evolved to photopolymerization printing of more intricate composites, while supercapacitor technologies that are less sensitive to solvents more often involved material jetting processes such as inkjet printing. In the last few years, the need to develop higher resolution printers, with multi-material printing capability has borne out in the performance data of batteries and other electrochemical energy storage devices. Underpinning every part of a 3D printable battery and many other devices is the printing method and the nature of the feed material. Material purity, printing fidelity, accuracy, complexity, and the ability to form conductive, ceramic, glassy, or solvent-stable plastics relies on the nature of the feed material or composite to such an extent, that the future of 3D printable batteries and electrochemical energy storage devices will depend on new materials and printing methods that are co-operatively informed by the requirements of the device and how it is fabricated. In this Perspective, we address the materials and methods requirements in 3D printable batteries and supercapacitors and outline requirements for the future of the field by linking existing performance limitations to the requirements of printable energy storage materials, casing materials and the direct printing of electrodes and electrolytes. We also look to the future by taking inspiration from new developments in additive manufacturing, to posit links between materials and printing methods that will allow new small form factor energy storage devices to be seamlessly integrated into the devices they power.**

# 1. Introduction

The 3D printing approach is a specific subset of additive manufacturing, where the materials are chosen and directly build from concept or design into a functional component. In the electrochemical energy storage scene, batteries and supercapacitors are dominant but typically come in a select number of form factors (shapes). The case for batteries is quite well known: cylindrical, prismatic cells, rectangular and coin cell etc. form the majority of Li-ion type products. These vary in size, but the form factor is well regulated. Similarly, for supercapacitors there are fixed form factors available. On rationale for 3D printing electrochemical energy storage devices (EESDs) is to circumvent the requirement for typical form factors, since 3D printing, like product design, is birthed by CAD on a computer. As such, designing the cell that stores and delivers power can in principle be done in parallel with product conceptualization and design. Microbatteries, ultracapacitors and ultra-thin film batteries have been developed and commercialized, partly with this end goal in mind, to provide a power source for appropriate products with minimal gravimetric and volumetric footprint. Much of the portable and consumer electronics, especially the Internet of Things (IoT), medical and personal healthcare devices will only require low power demands, and so limited size batteries and supercapacitors are ideal in this regards[1]. If they can be printed to seamlessly integrate into the product design, for aesthetic as well and comfort or functional reasons, the bulkier and fixed form factor standard battery need not be accommodated at product design stage.

3D printing and additive manufacturing (AM) in general, also provides an opportunity to form complex structures with ease compared to equally complex synthetic protocols and material assembly requirements[2]. And printing can in principle extend to metals, plastics, conducting composites, inorganic materials and fillers, polymeric ionically conducting electrolytes, and even bioinspired hierarchically structured composites[3]. 3D printing opens routes to rapid prototyping and fabrication that can also be massively parallelized, which also avoids the serial material roll-to-roll production currently used. Since the entire cell can be designed from the outset, 3D printing approaches can minimize the need for multiple fabrication steps but an innovation such as this requires advanced host materials and the assumption that a single 3D printing method is best for all components. While multi-material 3D printing has been portrayed as a possible leap in material printing solutions for EESDs which require several material formulations, there it may be the case that a multi-printer approach is needed to provide the optimum printing method for each active material or layer[4]. However, it is exciting that Lewis et al. have reported soft matter printing using multi-nozzle printing, thereby enabling massively parallel and complex multi-material printing at the voxel level[5]. Scheme 1 summarises some of the aims of AM in 3D printable batteries and supercapacitors, the materials, electrodes and printing innovation required to realise printable EESD technology.

The advantages of 3D printing and the overall AM approach to materials and battery cells[6] maintain the research interest and it is interesting to note the parallels being considered in the design aspect of 3D printed batteries and supercapacitors with those from the material chemistry community who for years have been incorporating synthetically prepared, often complex porous materials into EESDs. A pertinent example is the use of structured porous materials[7] using either metals or the active material themselves (anodes and cathodes). These structures are often periodically ordered, randomly porous, and range in pore lengths scales from a few nm to several microns in dimension. Maintaining electrical interconnectivity, reduced solid state diffusion limitations and reaction kinetics, while ensuring good capacity at repetitively high charging and discharging rates are goals that 3D printed cells would also wish to emulate. Many internal electrode designs incorporate complexity as it can be easily rendered on computer, with resolution in lithographic printers in the micrometer range. Advances in directly printing materials, especially soft and polymeric materials being deployed for soft robotics, shape-morphing systems and bio-inspired sensors[8]. These approaches bode well for polymeric electrolyte printing that buffers mechanical stress in electrode materials, or damage tolerant material printed directly[9].

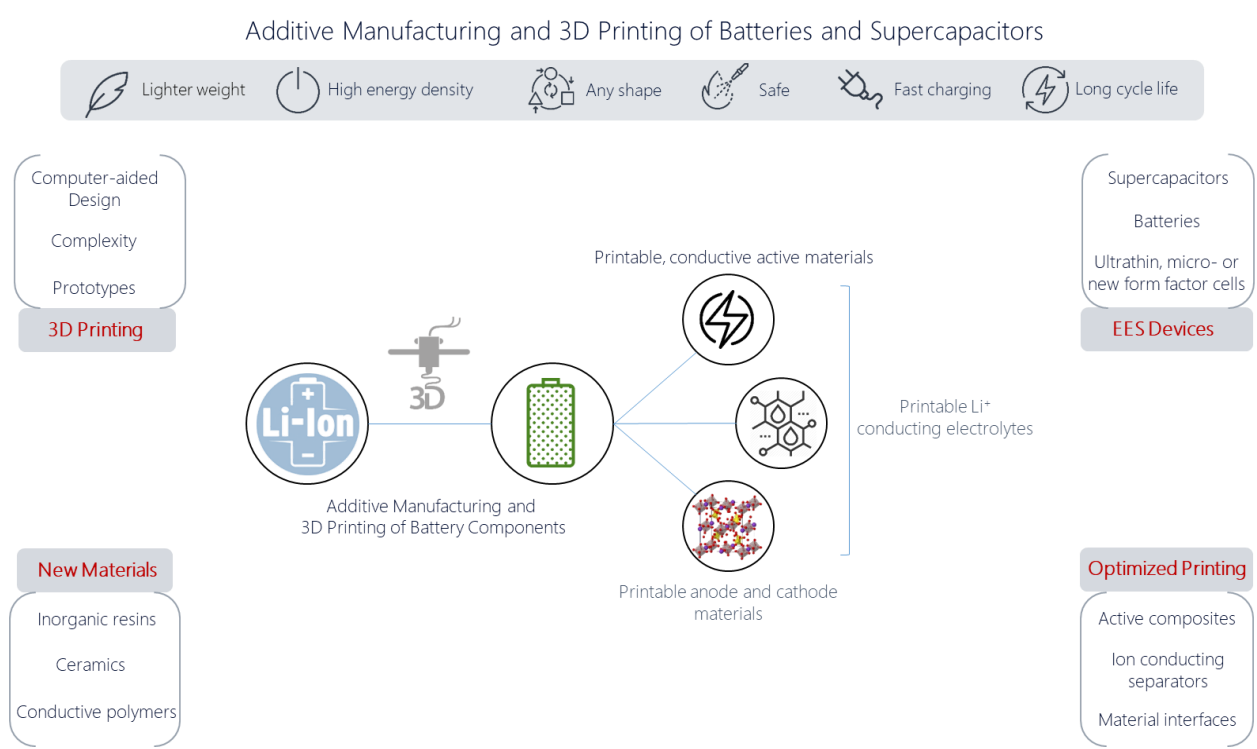

**Scheme 1** Basis and motivation for additive manufacturing and 3D printing materials, casings, electrodes, electrolytes and designs for electrochemical energy storage devices such as batteries and supercapacitors.

One application that requires alternative energy storage and delivery options is dense deployment of wireless (5G and beyond) telecoms nodes. Off-grid power sources are necessary for truly mobile, always-on connectivity. New forms of power sources with suitable energy densities and shapes to seamlessly form part of a wearable technology are also being urgently sought, for neo-natal care untethered and wireless, to energy autonomous healthcare technologies, smart home technologies, technology in the automotive industry, new consumer peripherals and integrated batteries for low-power sensors, food packaging and much more.

This Review focuses on AM approaches to 3D printing Li-ion batteries and supercapacitors, and avoids concepts related to AM that are more aligned with the 2D, 3D or porous materials community. To give the reader a relevant review of this field, while minimizing a repeating summary of previously published datasets from papers, we have carefully structured the content to link the 3D printing method to the end use material, by the properties and requirements of the printable material or composite. We begin by clarifying the various forms of 3D printing and look at the uptake of various printing methods in terms of pushed papers and filed patents around the world specifically for electrochemical energy storage devices such as batteries and supercapacitors, including some of the most recent advances. Using some of the initial reports from various 3D printed battery and supercapacitor systems, we showcase recent advances and then highlight future requirements for creating useful composites, electrolytes, separators, active materials, electrodes and casings using additive manufacturing or 3D printing methodologies to deal with limitation and trade-offs for printable EESDs. Importantly, we detail the limitations in host printing material as they stand for a range of 3D printing methods, which will need to be addressed for any real advance in this field. Finally, by contextualizing and comparing battery and supercapacitor reports, we deal with state of the art approaches that are tackling the primary limitation – the nature of the host materials required by many printing or AM methods, and how these challenges can be met to advance the viability of 3D printing batteries and supercapacitors with alternative form factors for future human-centric devices where the battery is considered at product design stage.

**2. 3D printing methods – state of the art**

Additive manufacturing (AM), commonly known as 3D printing, is becoming a powerful manufacturing strategy for fabricating functional 3D structures and driving the uptake of additive manufacturing into automotive and aerospace technologies, to a range of consumer products. Its popularity has grown from hobbyists to engineering professionals offering a remarkable control over designing novel architectures directly from computer aided design (CAD) software[10,11]. Contrary to traditional subtractive manufacturing where an object is carved from a bulk monolith, additive manufacturing allows complex structures to be printed layer-by-layer through a series of cross-sectional slices. (Figure 1a).

**Table 1:** List of ASTM categorized methods and specifications of various commercial products. See Figure 1 and associated caption for description of acronyms not defined in the main text/Table * Layer thickness values are estimated form several hobbyist and industrial products at the time or writing. We note that improvements in 3D printing are being announced frequently.

| ASTM Categories | Commercial Names | Build Platform (m) | Layer Thickness* ($\mu$m) | Build Rate | Materials (Possibility of Multimaterials) |
|---|---|---|---|---|---|
| *Material Jetting* | Inkjet | 0.5×0.4×0.2 | 16 | NA | Colloidal inks (Yes) |
|  | NPJ | 0.5×0.28×0.2 | 10-100 | 1.5 mm h$^{-1}$ | Colloidal Inks (Yes) |
| *Material Extrusion* | FDM, FFF | 1×1×1 | 100-800 | 500 cm$^3$ h$^{-1}$ | Plastic (Yes) |
|  | ADAM | 0.3×0.22×0.18 | 50-100 | NA | Metal (Yes) |
| *Binder Jetting* | 3DP | 0.4×0.25×0.25 | 30-200 | 3600 cm$^3$ h$^{-1}$ | Metals, ceramics plastic (Very Challenging) |
| *VAT Photopolymerization* | SLA, DLP | 0.8×0.33×0.4 | 10-150 | 22 mm h$^{-1}$ | Polymers, ceramics (Yes, Challenging) |
|  | CLIP | 0.3×0.3×0.3 | 25-100 | 5 mm h$^{-1}$ | Polymers, ceramics (Yes, Challenging) |
| *Powder Bed Fusion* | SLS, SLM, DMLS | 0.5×0.28×0.85 | 20-90 | 171 cm$^3$ h$^{-1}$ | Plastic, ceramics, metals (Yes, very challenging) |
| *Sheet Lamination* | LOM-DLP | 0.09×0.05×0.1 | 15-100 | N.A | Ceramics, metals |
| *Direct Energy Deposition* | DED, DMD | 1.2×0.8×0.8 | 800-1200 | 20 cm$^3$ h$^{-1}$ | Metals (Yes, Challenging) |

The thickness of each cross-sectional printed layer lies between 15-500 $\mu$m[1] depending on printing method and desired applications (Table 1). Currently, AM is limited to prototyping and custom-build small parts[11-13], however, to realize its use as a preferred industrial manufacturing tool key parameters such as layer thickness, build volume (the printable size of object), build speed (measured as the height of an object built in a given time (mm h$^{-1}$) or as a volumetric rate (mm$^3$ h$^{-1}$)), and materials properties need further improvement and optimization[14]. The last decade has witnessed numerous improvements in key parameters along with the development of various new technologies for additive manufacturing and printing-based methods. Therefore, the International Committee of the American Society for Testing and Materials (ASTM) has classified 3D printing methods into seven categories[15] including:

(1) Material extrusion (ME), more popularly referred to as fused filament fabrication or fused deposition modelling, and is widely available in relatively small sized standalone 3D printers (Figure 1a). Material extrusion also includes material dispensing (Figure 1b) which uses a liquid feedstock,

(2) Binder jetting (BJ) (Figure 1c),

(3) Material jetting (MJ) commonly known as inkjet printing (Figure 1d),

(4) Powder bed fusion (PBF) also known as selective laser melting or direct metal laser sintering (Figure 1e),

(5) VAT photo-polymerization (VAT-P) that includes stereolithographic apparatus, continuous liquid interface production (Figure 1f) and digital light processing,

(6) Sheet lamination (SL) is an additive manufacturing process in which sheets of material are bonded to form a part. Laminated object manufacturing (Figure 1g) is a related technique,

(7) Direct energy deposition (DED) comprises a range of terminologies such as laser engineered net shaping, directed light fabrication, direct metal deposition and 3D laser cladding (Figure 1h).

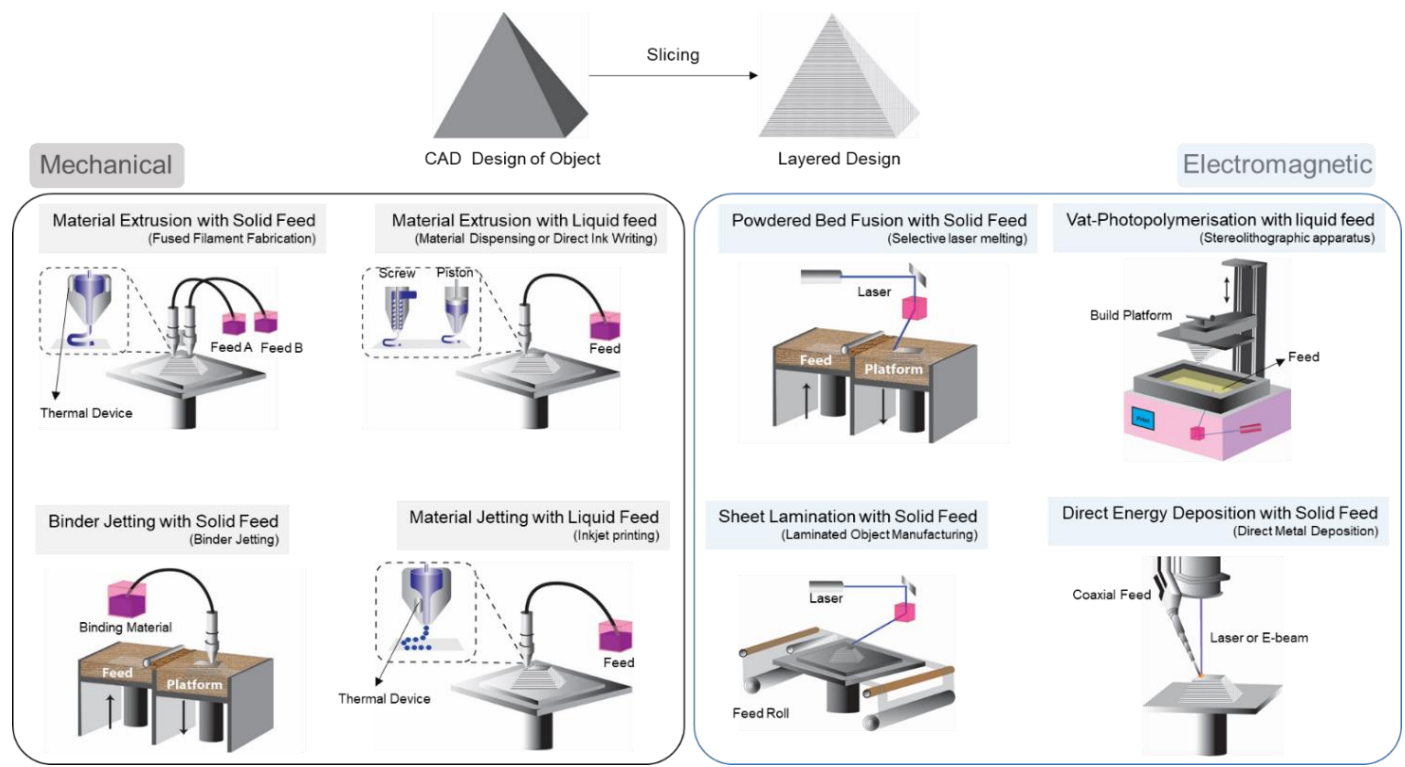

**Figure 1.** ASTM classified 3D-printing techniques separated into mechanical solidification methods and electromagnetically written printing. The primary design is generated by a 3D rendering software, followed by slicing into printable segments according to the printing method. (a,b) Materials extrusion (ME) and (c,d) Binder jetting/material jetting (BJ/MJ) use either solid or liquid-phase building materials as a feedstock. (e) Powder bed fusion (PBF), (f) vat polymerization (VAT-P), (g) sheet lamination (SL), and (h) direct energy deposition (DED) use either photon or electron optics and processes including photopolymerization, laser-induced heating, or laser cutting to generate 3D structures.

These technologies can also be classified by the physical state of building material (solid and liquid) or the method used to fuse the building material (mechanical or optical). For example, ME, MJ and BJ are a group of techniques where formation of 3D structures is based on mechanical forces whereas PBF, VAT-P, SL and DED use high power electromagnetic beams (light or electron beams) to construct a 3D object by melting, fusion or driving a chemical change resulting in solidification into a new shape at defined locations in space. Besides, MJ and VAT-P require a liquid phase build material (shown in blue in Figure 1) while most of the other techniques use solid powder or filaments as the building material.

To minimize the jargon of terminologies used in literature, all 3D printing methods here will be referred using ASTM terminologies. In passing, we remind the reader that ME techniques cover FFF or FDM based filament extrusion 3D printing, while VAT-P is more commonly referred to as SLA and DLP printing, and MJ broadly covers inkjet printing and related methods. Moreover, we have grouped these printing methods based on the physical state of the feed building material and the source used to fuse/construct a 3D object. For example, BJ and PBF are the two printing methods that use powder (metal or polymer) as building material. In PBF, a computer controlled laser or electron beam fuses the powdered material (Figure 1e) whereas BJ utilizes a liquid binder through multiple inkjet nozzles[16] (Figure 1c). Once a layer of powder is bonded or fused together, a new layer of powder is spread using a roller and the process of fusion is repeated. A key advantage of both techniques lies in the use of a powder bed which not only provides the building material but serves as an in-process support allowing complex shapes with high geometrical accuracy[17]. These considerations are important when designing active materials for energy storage devices that themselves are printed into uncommon form factors. Nevertheless, porosity is common in these powder bed processes requiring hot isostatic pressing or infilling with another material to improve the mechanical properties of the finished product[18,19]. To fabricate a dense and mechanically stable 3D structures such as active material constructors or current collectors, DED is another powder- or wire-based printing method that utilizes a laser or electron beam focused on a substrate producing a melt pool to which a coaxial powder stream or a wire feed is injected, building a 3D structure[20,21] (Figure 1h). Despite producing robust structures, the process is time consuming and requires inert conditions making it expensive for industrial use. A lower cost option is to use a material extrusion (ME) process such as fused filament fabrication where material is drawn through a computer controlled nozzle where it is melted and deposited layer by layer to a desired shape[22] (Figure 1a). The process uses readily available ABS and PLA thermoplastics, however metal particles coated in plastics have been recently introduced[23]. Another widely used method of material extrusion involves liquid feed being dispensed though a screw or piston based nozzles with diameter ranging from 0.1-250 μm[24,25] (Figure 1b). As the direct ink writing (DIW) method uses liquid feed, it eliminates the need for any heating device making it simple and cost effective. Material jetting (MJ) is another liquid feed printing method where multi-material inks can be used to fabricate structures making it a common printer for hobbyists and researchers. Inkjet printing approaches are not 3D printers, but a laminate ink printing approach with some degree of thickness control. The method uses liquid ink which is jetted on to build surface where it solidifies to form desired shapes (Figure 1c,d). The printing capabilities and quality of the finished product are dictated by the Ohnesorge number $Z = \sqrt{\rho \sigma d}/\mu$, a quantity that depends on viscosity (μ), surface tension (σ) and density (ρ) of the ink. Droplets are controlled using thermal or piezoelectric actuators and ink compositions with $1 < Z > 10$ are expected to produce stable droplets[26-28]. Recently, magnetojet actuators have been developed for depositing molten metal[8] which was not possible with traditional thermal and piezoelectric actuators.

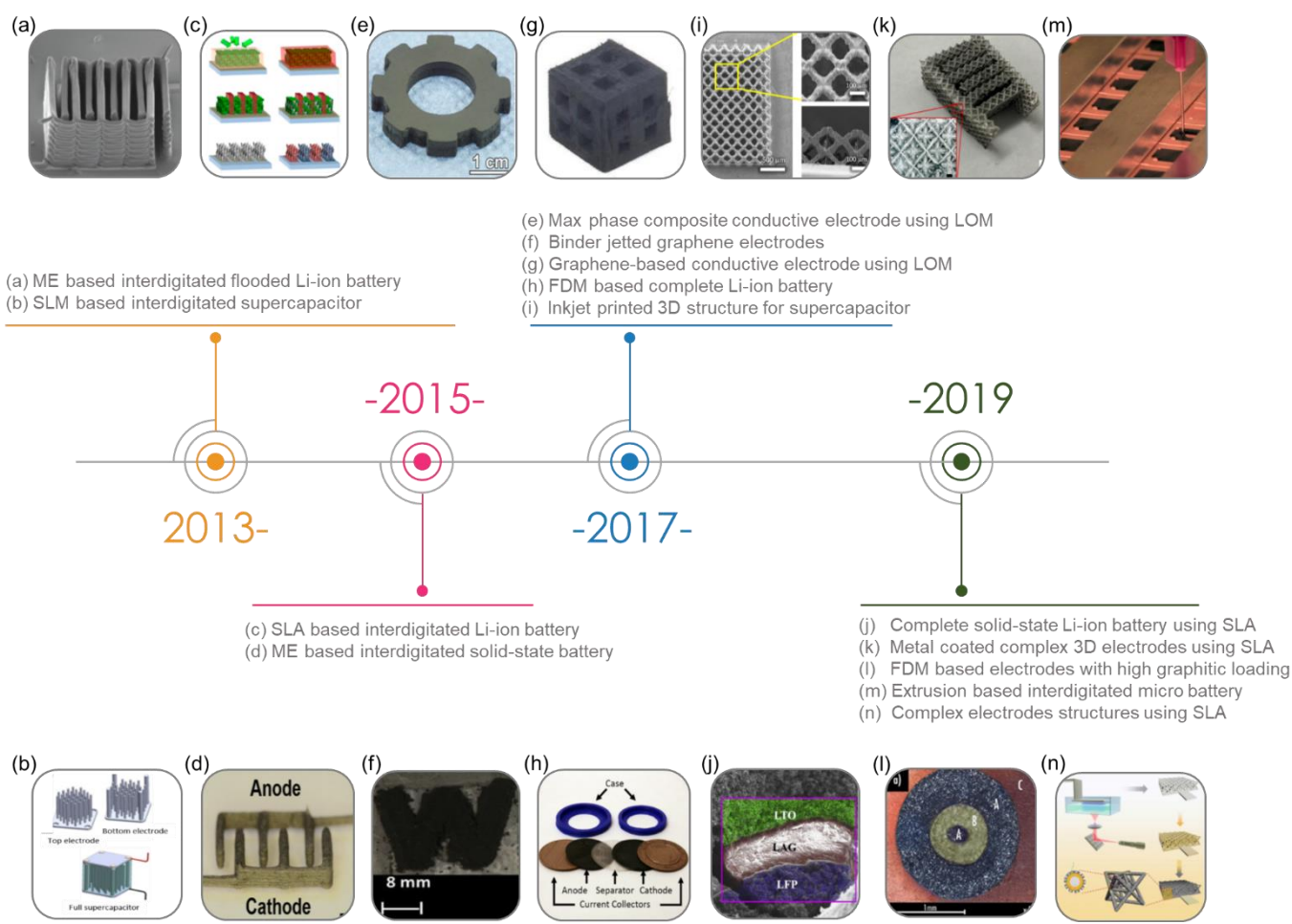

**Figure 2.** A timeline of 3D printing methods, materials and structures for energy storage devices from 2013 through 2019. Representative structures from several published reports (a-n) are adapted from Refs [29-41], with the permission of Royal Society of Chemistry and American Chemical Society.

One of the fundamental problems with material extrusion and jetting techniques is the rough surface of the final product requiring additional finishing step[36,42]. To obtain high resolution smooth surfaces, vat-photopolymerization (VAT-P) is another printing technique that uses selective photo-polymerization of polymer resin to build a layer-by-layer 3D structure. Once a polymer layer is cured by a light source, another layer is allowed to be formed on the surface and the process of curing is repeated[43]. Unfortunately, the fabrication process in traditional VAT-P is very slow, therefore, many variants have been developed including digital light processing (DLP) and continuous liquid interface production (CLIP)[44]. Finally, SL or laminated object manufacturing is a novel 3D printing method involving bonding/fusing of multiple metal or plastic laminates which are sliced using a knife or laser cutter. The applied laminates, including paper, plastic, and metals, indicate the potential for fabrication and packaging of customized sandwich-type electrochemical devices[33,35,45,46].

All the techniques discussed above have their advantages and disadvantages in terms of resolution, build speed, build volume and multi-material printing capability. Nevertheless, tremendous amount of research is being dedicated to developing new ways of additive manufacturing while advancing existing technologies. Figure 2 maps the timeline of various energy storage devices developed in last six years. Although most of the work has been focused on material extrusion and jetting based methods, however, some interesting electrode structures have been proposed using other 3D printing method such as VAT-P, MJ and PBF. We also performed an extensive literature survey of publications and patents for 3D printing technologies and their application to energy storage materials and devices. Figure 3a shows the number of publications related to each of the classified methods from Figure 1, while Figures 3b,c highlights the amount of research published in three different geographical regions (Europe, USA, and Asia) in total, and the number of those reports specifically related to energy storage, respectively. Finally, Figure 3d provides a perspective on the advantage and disadvantages of each printing method based on five key parameters. We believe the information is quite useful for selecting a 3D printing method for realizing commercial production of 3D printed lithium ion batteries.

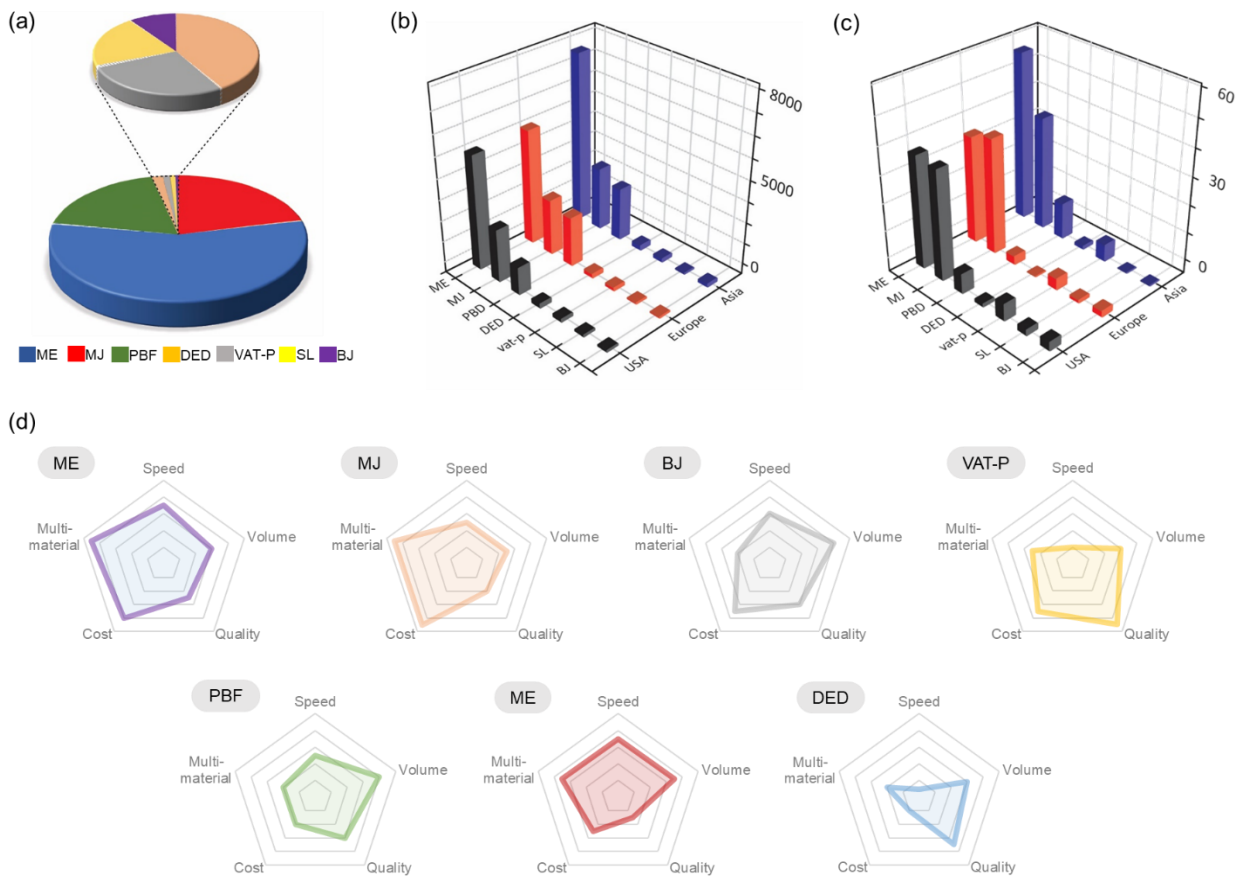

**Figure 3.** Pie charts of (a) number of publications from 2000-2019 related to each 3D printing method. The data was acquired on 25[th] September 2019 using the Scopus database. The following queries were searched in the article title, abstract and keywords of publications. For MJ: (Material jetting) OR (Inkjet Printing), for ME: (Direct ink writing) OR (Fused deposition modelling) OR (fused filament fabrication) OR (Direct ink writing), for VAT-P; (VAT Photopolymerization) OR (stereolithography apparatus) OR (digital light processing printing), for BJ: (Binder jetting printing), for SL: (Sheet lamination printing) OR (laminated object manufacturing AND 3D printing), for PBF: (powdered bed fusion) OR (Selective laser melting) OR (Selective laser sintering), OR (Direct metal laser sintering), for DED: (directed light fabrication) OR (Direct metal deposition printing) OR (3D laser Cladding). All search words were carefully selected to obtain results related to additive manufacturing or 3D printing. (b) All publications related to each technique and geographical region. (c) All publications related to batteries and supercapacitors. (d) Radar charts grading the most important parameters in 3D printing technology on a scale where 0 = lowest suitability and 9 = highest suitability. In (a) and (b), the suffixed -ES to each acronym relates to energy storage papers/patents using that technique.

## 3. 3D Printable Materials and Composites

There are numerous reports in the literature on 3D printing various parts of electrochemical energy storage devices (EESDs) and even whole devices. Most EESDs are composed of structural (cell casing, current collector, separator) and functional (electrolyte, anode and cathode) components[47]. In this section, an overview of additive manufacturing examples of EESD components and materials used for their fabrication by different 3D printing methods is given. We provide insights into the optimal choice of 3D printing technique for different material types and applications along with the issues that should be considered.

### 3.1 Conductive Materials and Cell Casing

3D printing of EES device casing and current collectors is a relatively less challenging task compared to direct printing of active materials in pure form or as a composite, due to the wide variety of suitable raw materials for many printing methods.

*Cell casing/enclosure* can potentially be 3D printed using almost any available printing technique and variety of materials. However, it is important to consider the design strategy and the type of EESD that will be enclosed in the casing. For example, in the fabrication of Li-ion batteries it is essential that the final assembly is airtight to prevent atmospheric gases and moisture from entering the device interior, and to eliminate the leakage of electrolyte. In this case a 3D printing method capable of producing high-precision, non-porous solid objects, for instance stereolithography (VAT-P), is preferred. The operating conditions (e.g. temperature range, environment) and application requirements (e.g. flexibility, mechanical stability, light weight) can also dictate the choice of casing material and 3D printing method.

*Current collectors* are an important component of EESD in most cases where the electronic conductivity of active material is insufficient. Therefore, the integration of current collectors in the EES device design is generally necessary for fabrication of high-performance 3D printed devices. Besides, the feasibility of additive manufacturing of highly conductive composites is important for the active material preparation and can eliminate the need for a separate current collector.

Although *metals* are the conductive materials of choice in traditional EESD manufacturing process, their application in additive manufacturing of fully printed devices is limited mostly due to extreme printing or post-treatment conditions and relatively high cost of equipment. Nonetheless, there are examples of 3D printing of standalone metal 3D structures used as current collectors and scaffolds for active material deposition. These approaches offer exceptional design versatility and variety of 3D shapes, high surface area and electronic conductivity, enhanced control over electrode morphology and electrodeposited layer thickness, increased contact area between current collector and active material for hierarchical 3D structures, and high mechanical strength that can counter the volume expansion of deposited active materials[48-50]. Metallization of stereolithographically 3D printed polymeric microlattices was used for preparation of conductive hierarchical templates for subsequent active material deposition[51,52].

Another way to use metals as conductive materials in 3D printing is the preparation of dispersions and polymer composites containing metal powder or nanoparticles (typically Cu, Ag, Au, Al)[53-56]. In this case the optimal content of metal powder in the polymer matrix must be found in order to exceed the percolation threshold, and, at the same time, to preserve the material printability. Very high loading of metal particles usually makes composite filaments brittle due to formation of agglomerates and voids in the polymer matrix. Overall anisotropy in inhomogeneity of the printed structures affects the material printability and mechanical strength[54]. Nonetheless this approach does not involve extreme processing conditions and therefore can be integrated into multi-material 3D printing of EES devices. Using liquid or low-melting metal alloys (typically, gallium-based) and their mixtures with nanomaterials is a different route to fabrication of stretchable, flexible and highly conductive 3D objects using additive manufacturing process under mild conditions. Several studies report on extrusion-based techniques for fabrication of 3D printed liquid metal structures[57-59].

Initially developed as a two-dimensional printing method, electrohydrodynamic printing (EHDP) has now been evolving to a 3D additive manufacturing process. EHDP, sometimes called as electrohydrodynamic jet printing or e-jet printing, is based on electrohydrodynamically induced ink flow (in the form of jets or droplets) from a nozzle, which is somewhat similar to the traditional MJ techniques[60-63]. As an example, recently an ink-free additive manufacturing technique called electrohydrodynamic redox printing (EHD-RP) has been developed and applied for fabrication of multi-metal 3D structures[64]. This method is based on electrochemical dissolution and re-deposition of metals; it enables direct fabrication of multi-metal structures with resolution as low as 250 nm and feature size less than 400 nm without the need for post-processing. Compared to the traditional 3D printing methods, EHD-RP offers exceptionally high resolution, good multi-material printing capability, precise control of the local chemical structure and morphology.

*Carbonaceous materials and composites* are another class of electronically conductive materials that have been extensively studied in recent years. Graphite, graphene, graphene oxide (GO), reduced graphene oxide (rGO), carbon nanotubes (CNT), carbon blacks and active carbons are widely used carbon-based materials for preparation of 3D printable conductive formulations for EESD[65,66]. The intrinsic advantages of these materials are high electronic conductivity, high surface area and porosity, good chemical end electrochemical stability. High lithium intercalation capability of many carbon forms[67-72] can also be beneficial for Li-ion battery fabrication.

*Graphene* has a unique combination of outstanding electron mobility, high specific surface area, high intrinsic specific capacitance, good chemical stability, flexibility, optical transparency, exceptional mechanical strength, which makes graphene a promising candidate for EES applications[73-77]. Graphene oxide can be viewed as graphene functionalized with various oxygen-containing groups (epoxide, carbonyl, carboxyl, hydroxyl). Conductivity of GO depends on the C/O ratio, and in a reduced form (rGO) can be as high as $10^4$ S cm$^{-1}$ [78,79]. As opposed to graphene, CNT, active carbons and carbon blacks, graphene oxide is amphiphilic and can form stable aqueous colloid solutions as well as dispersions in other common solvents, which facilitates the formulation of 3D printable inks with necessary rheological properties for extrusion-based printing methods without addition of polymer binders[80]. 3D printed low-conductivity GO can be subsequently reduced to form highly conductive rGO patterns.

*Graphene-based materials* have been employed in conjunction with various 3D printing methods (MJ, ME, VAT-P, PBF and their variations[81]) to create conductive structures, patterns and aerogels, which can be used as current collectors, electrodes or templates for electrodeposition. 3D printed graphene composites typically demonstrate high surface area and porosity, good mechanical strength, high conductivity and capacitance, excellent electrochemical and chemical stability[51,74,82-92]. A detailed review focused primarily on the application of graphene-based materials in additive manufacturing of EES was published by Fu *et al.*[65] Besides graphene-based materials, a novel class of electronically conductive 2D nanomaterials called *MXenes*, 2D transition metal carbides and carbonitrides, have been attracting a growing attention in recent years. These materials show great promise in additive manufacturing of EESD as current collectors, supercapacitor electrodes and active material components[93-95] due to their high capacitance, high electric conductivity and superior charge storage and transfer capabilities[96]. Similar to graphene oxide, hydrophilicity of MXenes allows them to be easily dispersed into aqueous colloids, which facilitates preparation of 3D printable compositions[93].

*Active carbon*, *carbon blacks* (amorphous carbon) and *graphite* are widely used materials for preparation of conductive 3D printable compositions due to their electric conductivity, low cost, simplicity of handling and production, chemical and electrochemical stability, high porosity (especially for active carbons). Some of these materials are also capable of reversible Li ion intercalation and have appreciable intrinsic specific capacitance[72,97]. They have been utilized in various 3D printing techniques (extrusion-based, IJP, SLS) in the form of polymer-based conductive composites used for fabrication of electrically conductive structures, supercapacitor electrodes, Li-ion battery electrodes[97-102].

*Carbon nanotubes* are promising candidates for current collector and electrode 3D printing due to high carrier mobility, superior mechanical strength and large specific surface area that can be functionalized for improved energy storage performance[103,104]. Depending on the structure, CNT can be either conductors or semiconductors. However, the relatively high cost of carbon nanotubes production can be a limiting factor for wide adoption of CNT-based materials in EESD manufacturing. To achieve good dispersion of CNTs in common solvents and prevent aggregation, various polymers and/or surfactants are often added to the solvent[105,106] or, alternatively, chemical modification of the nanotubes is performed, for example, carboxylation of CNT, which increases hydrophilicity and allows a preparation of aqueous dispersions[107]. This decreases the cost of ink preparation and provides easy handling and storage as well as VOC-free environmentally friendly processing.

Extrusion based 3D printing of polymer/CNT composite allowed the fabrication of conductive features as small as 100 μm, exhibiting good electrical conductivities (up to 100 S m$^{-1}$)[83,108]. CNT containing compositions were used in 3D printing of supercapacitor electrodes[109,110], mechanically reinforced liquid metal wires for flexible electronics[59] and various freestanding conductive 3D microarchitectures[111] with conductivities up to ca. 2500 S m$^{-1}$.

A technique (C-MEMS), in which polymeric photoresist patterns are pyrolyzed to carbon was developed by Wang et al.[112]. Similar methods can be developed for high resolution 3D printed polymer structures (e.g. by using VAT-P) to fabricate hierarchical porous conductive carbon-containing materials for EES devices[113].

The considerations previously mentioned for printability of metal-based composites hold true for carbonaceous composites. Conductive agent percolation threshold, brittleness of composites at high carbon material loading and material anisotropy should be taken into account during the composite formulation. It is also important to account for the possible printer nozzle wear when using such abrasive conductive fillers as graphene, carbon nanotubes, metal powders and some other[83].

Structural anisotropy is an intrinsic characteristic of parts printed using extrusion-based methods[114], meaning that the structures in horizontal direction and vertical direction of printed parts are substantially different. Structural anisotropy of the conductive composite prints naturally entails the anisotropy in resistivity[98,99]. More specifically, due to the layered structure of the printed parts, the continuous conductive path lies mostly in the horizontal direction of the fibers. In the perpendicular direction, the formation of a conductive path strongly depends on the fusion between adjacent layers, which is influenced by the printing parameters. For similar reasons, the printed structure's orientation during the extrusion printing can influence the electrochemical behavior of printed composites with vertical orientation being more favorable[102]. As opposed to extrusion-based techniques, other methods such as VAT-P, PBF or MJ can render significantly less anisotropic prints both in horizontal and vertical directions due to the uniform layer building process and more compact layer stacking.

## 3.2 Polymer and Solid Electrolytes, Separators

The electrolyte in EESD serves as a medium for ion transfer, storage, and electrode separation and has a great influence on electrochemical characteristics such as rate capability, voltage range, cycling performance[47,115]. The key parameters of electrolytes include ionic conductivity, electrochemical window, stability, safety and operating temperature range. For obvious reasons, it is not practically possible to print three-dimensional structures with low viscosity liquid electrolytes, which can only be drop casted or injected in already printed devices. Therefore, only solid-state or gel electrolytes can be 3D printed (without post printing jellification or solidification), which have such important advantages as relatively high ionic conductivity, nonflammability, improved thermal and chemical stability, no leakage issues, possibility of integration into all-3D-printed designs[116,117]. Additionally, solid or gel electrolytes can in many cases replace separators, which reduces complexity in EES device designs.

*Gel electrolytes* can be classified as aqueous, organic, ionic liquid-based and redox-active gel polymer electrolytes[118]. Gel polymer electrolytes (GPEs) typically consist of a host polymer matrix (polyvinyl alcohol, PVA; polyethylene glycol, PEG; polyethylene oxide, PEO; polyacrylonitrile, PAN; poly(methyl methacrylate) PMMA; poly(vinyilidenefluotide), PVDF, and its various copolymers, e.g. with hexafluoropropylene, HFP; functionalized cellulose; and some others), inorganic ionogen (a compound producing ions when dissolved, e.g. salt, strong acid or base), solvent (water or organic solvent) and sometimes plasticizers and inorganic fillers to improve mechanical, thermal and conducting properties[118,119]. Several criteria that define the suitability of different components of GPE can be formulated. A good polymer matrix should have wide electrochemical window, low glass transition temperature, high molecular weight, good thermal stability, functionalities and structure that can facilitate ion transport. A suitable ionogenic compound ideally exhibits: full dissociation with minimal ion aggregation; high thermal, chemical and electrochemical stability; high solubility in the chosen solvent; high ion mobility. A proper solvent should have a combination of high dielectric constant, low viscosity and high electrochemical, chemical and thermal stability.

Among the gel electrolytes, *aqueous gel polymer electrolytes* are highly attractive due to their high ionic conductivity, low cost, environmental friendliness, safety and processability[120]. However, aqueous electrolytes naturally have narrow electrochemical windows due to water splitting, which limits their areas of applicability. For example, they are not ideal for Li / Li-ion batteries and some other high voltage systems. The choice of the electrolyte ionogen strongly depends on the electrode materials and application (SC, aqueous battery, etc.). Typical examples of inorganic ionic compounds used for the preparation of aqueous gel electrolytes include neutral salts (e.g. LiCl, $LiClO_4$, $Na_2SO_4$), strong acids (e.g. $H_2SO_4$, $H_3PO_4$) and strong bases (e.g. KOH, NaOH, LiOH). Notably, the ionic compounds used for the preparation of aqueous GPEs must not undergo hydrolysis in order to achieve higher conductivity and to avoid undesirable side reactions and electrolyte degradation. It is also worthy to mention that acidic or alkaline solutions can cause corrosion of metallic dispensing nozzles and damage other parts of the printing system, which are in contact with such electrolytes. Therefore, all possible interactions of aqueous gel electrolytes with the printer building materials should be carefully considered prior to 3D printing[117].

The issue of narrow operating voltage window of aqueous gel electrolytes can be addressed by using *organic gel polymer electrolytes.* The composition and preparation method of organic GPE greatly affect the ionic conductivity and mechanical properties of the system. Generally, organic gel polymer electrolyte inks are prepared by mixing a polymer with high molecular weight, e.g. PMMA, PVDF, PVDF–HFP, with a conducting salt (e.g. $LiPF_6$, $LiBF_4$, LiTFSI, $LiClO_4$) in a nonaqueous solvent. Commonly used organic solvents are carbonates (dimethyl carbonate, DMC; ethylene carbonate, EC; fluoroethylene carbonate, FEC; ethyl methyl carbonate, EMC; propylene carbonate, PC), dimethyl formamide (DMF), dimethyl sulfoxide (DMSO), tetrahydrofuran (THF), various fluorinated solvents and the mixtures thereof[116,118,121]. With the use of organic GPEs, the operating voltage can be extended beyond 3.5 V, which is suitable for Li-ion battery applications. The increase of the cell voltage window is also beneficial for improving the energy density of other EESDs. Organic GPEs are generally much less aggressive towards the metallic components in 3D printers.

Compared to aqueous and organic electrolytes, *ionic liquid-based gel electrolytes* show several extra advantages, such as nonvolatility, nonflammability, wider operating voltage windows, thermal and electrochemical stability[122,123]. Ionic liquids (ILs) are low-melting salts, which commonly have such attractive properties as negligible vapor pressure, ionic nature and hence ionic conductivity, wide liquid range, wide electrochemical window and good thermal stability. Many IL-based GPEs are even more suitable for 3D printing of EESDs than aqueous and organic GPEs due to high stability and lack of corrosivity to metallic and plastic parts of the printer. Ionic liquids can replace commonly used organic solvents and water in GPEs serving as a solvent, salt and plasticizer at the same time. However, due to relatively high viscosity of ionic liquids and ion-pair formation, the addition of inorganic salt and/or diluent may be necessary in order to increase the charge mobility or to add the required ions (e.g. in the case of Li-ion batteries)[124]. The 3D printed IL-based gel electrolytes demonstrate superior thermal stability, a high ionic conductivity up to several mS cm$^{-1}$ at ambient temperature and good mechanical flexibility.

*Redox-active* GPE is a gel polymer electrolyte containing a small amount of redox-active additives, which can significantly increase the specific capacitance of supercapacitors. In this case additional electrochemical redox reactions contribute pseudocapacitance to the overall capacitance, thus gaining capacitance not only from electrode materials but also from the electrolyte. Redox-active species such as organic molecules (hydroquinone, methylene blue, indigo carmine, p-phenylenediamine, m-phenylenediamine, lignosulfonates) and inorganic compounds (e.g. $K_3Fe(CN)_6$, KI, $VOSO_4$, $Na_2MO_4$, $CuCl_2$) have been studied. The GPEs containing redox-active mediators have been explored in supercapacitors, pseudocapacitors and Li oxygen batteries[117,118].

Due to the rheological properties of gel electrolytes, extrusion-based 3D printing techniques are mainly used for their fabrication. *Solid-state polymer and ceramic electrolytes* are comparatively less common in additive manufacturing of EESD, but wider range of printing methods (including ME, VAT-P and MJ) can potentially be utilized in this case. Solid state electrolytes have a number of advantages over liquid and gel electrolytes, such as higher mechanical and thermal stability, wider electrochemical windows and improved safety[116,125,126]. However, it might be challenging to integrate solid state electrolytes in continuous manufacturing processes, because the same 3D printing technique is desirable for all the device components. Good contact between the electrode materials and electrolyte is essential in order to reduce the interfacial resistance of the system, and 3D printing methods have a potential to improve this contact. The low interfacial resistance is highly important for achieving good rate and cycling performances of EESD, especially since the bulk conductivity of solid-state electrolytes is generally lower than the conductivity of liquid or gel electrolytes[126]. The 3D fabrication of solid electrolytes for Li-ion batteries containing lithium conductive ceramics has been reported[125,127]. Electrophoretic deposition of solid-state ceramic electrolytes on 3D printed patterned substrates is another possible method, which can be used not only for electrolytes, but also for electrode materials[52].

As mentioned earlier, in many cases where a suitable 3D printed solid state or polymer electrolyte is present, it is not necessary to include a separator in the EESD design. Nevertheless, for the systems containing liquid electrolyte the use of separator can be essential for improved stability and to prevent short circuiting. Several examples of additive manufacturing of porous separator structures using a variety of 3D printing techniques have been published[128-131]. Moreover, 3D printing can make it possible to improve EESD performance via fabrication of separators with desired structural, thermal, mechanical and chemical properties that can be achieved by adding fillers with required characteristics, for example metal oxide nanoparticles, boron nitride and so on[130,131].

### 3.3 3D Printed Active Materials

Currently, the research in the field of electrochemical energy storage is mainly focused on two types of EESDs: supercapacitors and batteries (e.g. Li-ion, Li – $O_2$, Li – S, Zn-air). Electrodes are arguably the most critical components of EESD as their structure, properties and composition largely define the storage characteristics and cycling performance of the device. The performance of EESDs can be improved by using hierarchically structured porous electrodes with interconnected nano- and microscale pores, which can provide shorter diffusion and ion-transport pathways, increased surface area and surface availability[132]. The engineering and fabrication of such electrodes could be relatively more straightforward with the aid of 3D printing methods than with conventional techniques. As opposed to planar designs, various 3D electrode architectures are able to promote more efficient utilization of available device volume having the same footprint as planar counterparts thus increasing specific volumetric and areal energy densities[104].

The two most common electrode arrangement designs used in 3D printing of EESDs are classic sandwich-type configuration and in-plane configuration[133]. In a sandwich-type arrangement, each component of the device is placed in a separate layer, and all these layers are stacked in the final device. In-plane configuration combines all the components of EESD into interdigitated or alternating structures, which are placed on one plane. For this approach to be successfully implemented, the 3D printing method should be able to perform multi-material printing starting at the same vertical axis level for all the printed components. Therefore, the techniques, which are based on direct material deposition (e.g. ME, MJ), are preferable for the fabrication of in-plane configurations. The in-plane 3D printed designs have multiple advantages such as possibility of better miniaturization and integration with other microelectronic devices, increased contact area between EESD components, shorter diffusion paths, and more efficient utilization of available surface and volume.

Electrode materials used in *supercapacitors* typically include carbonaceous materials, metal oxides, conductive polymers, novel 2D nanomaterials (e.g. MXenes and black phosphorus), metal organic frameworks (MOFs) and metal nanoparticles[134], which can be either electric double layer (e.g. carbon materials) or pseudocapacitive materials (commonly, metal oxides and conductive polymers) depending on their charge storage mechanism. The 3D printing of MXene, carbonaceous and metal composites has already been discussed in section 3.1.

Common *metal oxides* for SC electrodes include NiO, CoO, $RuO_2$, $MnO_2$, $NiCo_2O_4$ and $V_2O_5$[117]. However, metal oxides have poor electrical conductivity and tend to degrade during cycling. Hence, binders and conductive fillers are often required for preparation of metal oxide inks.

*Conductive polymers* (CP) have considerably higher electrical conductivity than metal oxides, thus they can be used as electrodes without conductive additives and current collector. They also demonstrate mechanical flexibility, easy dispersion in solvents and relatively low cost. Similar to many other materials for EESDs, CPs can be electrodeposited on conductive substrates[50], which is useful for fabrication of hierarchical patterned electrodes. The widely used conductive polymers are polythiophene (PTH), polypyrrole (PPy) and polyaniline (PANI)[135]. They can be easily added to printable inks, which makes CPs a promising class of materials for additive manufacturing of SCs.

*Metal – organic frameworks (MOFs)*, materials constructed from metal-containing nodes and organic linkers, are a promising group of electrode materials due to high porosity, controlled pore size, structural diversity and chemical stability[136]. For example, FDM printed ABS – MOF composites were obtained by Kreider et. al.[137]. Extrusion-based 3D printing method was used to fabricate Co-MOF-derived porous cathode for Li–$O_2$ batteries[138]. *Covalent organic frameworks (COFs)* are highly porous structures comprised of organic building units connected by strong covalent bonds (B–O, C–N, B–N, B–O–Si)[139]. Much like MOFs, COFs show great promise for EES applications due to controllable pore sizes, high surface area and design diversity[140].

The structure and composition of electrode materials for *batteries* depends on the type of battery (Li-ion, Na-ion, Li – $O_2$, Li – S, Zn-air, etc.) and the type of electrode (negative or positive). 3D printing of positive electrode (cathode) materials (and even full cells) for metal-air and lithium-sulfur batteries has been reported[123,138,141,142]. In the present review we consider mainly 3D printable electrode materials for lithium-ion batteries. The *negative electrode (anode)* in Li-ion batteries typically consists of an active material, which is capable of reversible lithiation/delithiation during battery charge / discharge cycling, a polymer binder and conductive additive (where the electronic conductivity of the anode composition is insufficient)[115]. Most frequently used anode active materials in 3D printed Li-ion batteries are graphite, graphene and lithium titanate ($Li_4Ti_5O_{12}$, LTO)[52,64,85,100,143-145]. Polymer binders and matrices used for preparation of printable compositions can be thermoplastic polymers (e.g. polylactic acid, PLA; acrylonitrile butadiene styrene, ABS), functionalized cellulose, PVDF or aqueous GO[85,100,143,146,147].

A representative work describing the preparation of highly loaded (up to 62.5 wt %) graphite – PLA conductive composites used as filaments for FDM 3D printing of Li-ion battery negative electrodes was published by Maurel et al.[100] Such a high graphite content naturally caused brittleness of produced samples and required the addition of plasticizers (at least 20 wt % and less than 60 wt %) in order to be printable by FDM method. The highest achieved conductivity was ca. 0.2 S cm$^{-1}$ (without conductive additives) and ca. 0.4 S cm$^{-1}$ (with Super-P carbon black). High graphite loading made it possible to achieve quite considerable specific capacities of 200 mAh g$^{-1}$ at C/20 and 140 mAh g$^{-1}$ at C/10. That said, at higher current densities the specific capacity significantly decreased.

The working principle of lithium-ion battery *positive electrode (cathode)* is in a sense similar to the anode, with the difference in the direction of lithium ion flux during charge/discharge cycling. In most cases active materials used in LIB cathodes are mixed lithium – transition metal oxides and phosphates, such as lithium cobalt oxide $LiCoO_2$ (LCO), lithium manganese oxide $LiMn_2O_4$ (LMO), lithium nickel manganese cobalt oxides $Li_xNi_yMn_zCo_{1-y-z}O_2$ (NMC), lithium nickel cobalt aluminum oxide $LiNiCoAlO_2$ (NCA), lithium iron phosphate $LiFePO_4$ (LFP) and lithium manganese iron phosphate $LiMn_xFe_{1-x}PO_4$ (LMFP)[47]. The cathode active materials used in the additive manufacturing of Li-ion batteries include LFP, LMFP, LCO and LMO, lithium iron phosphate being the most widely studied (perhaps due to its high stability and excellent processability). As in the case of anode materials, polymer binders and conductive additives can also be included in the printable composition to achieve the required conductivity, mechanical and rheological properties of the cathode material[52,64,143-150].

Most of the considerations previously discussed for 3D printing of metal oxides and LIB anode materials are also true for LIB cathode materials. The difference between theoretical capacity of active materials and the obtained values along with the poor C-rate performance, commonly reported in connection with polymer-based composites and material extrusion 3D printing techniques, can be explained by insufficient electronic and ionic conductivity of the printed materials due to complete isolation of some amount of conductive materials within polymer matrix. Possible approaches to address this issue could be further boosting of conductive material loading, increasing polymer matrix porosity (for example, by chemical pre-treatment[85]), designing more complex scaffold structures for 3D printed electrodes, or using 3D printing methods and material formulations that do not involve an insulating polymer matrix. We will discuss these general requirements for a good 3D printable EES host electrode material further on.

Increasing printed electrode surface area can however have a negative effect on the cell performance due to the higher exposure to electrolyte. During battery cycling, the formation of the SEI layer on the electrode surface consumes electrolyte and lithium resulting in low initial Coulombic efficiency and can significantly reduce battery capacity and energy density[47,115]. In tandem, solvent decomposition can degrade the fidelity and shape of the printed electrode or structure, and cause dewetting phenomena between the host and any inclusion or filler materials, adversely affecting its efficacy as a functional electrode. Using solid or gel electrolytes, optimization of electrolyte composition, electrolyte additives, and developing electrodes with ion-conductive protection layer can help to solve these problems.

Based on our analysis of literature reports, the majority of research reports using 3D printing of electrode materials for EESDs implement different variations of ME and MJ-based printing techniques. This fact can be explained by the simplicity and relatively low cost of equipment, multi-material printing capability, facile fabrication of printable material composition, porosity of the printed structures and relatively high speed of these methods.

## 3.4 Characteristics of Useful 3D Printed Materials

During the 3D printing process, the source material properties and preparation as well as printing method limitations are the key considerations to achieve the desired performance and functionally useful and stable structural characteristics. Different 3D printing techniques have their own combination of advantages and limitations, which should be considered for each particular application and energy material. The most important parameters of 3D printing method related to the printed material performance and EES device fabrication are resolution, printed material porosity, multi-material printing capability, source material requirements (e.g. rheology, composition), physical processes involved in printing, and printing speed. In Figure 4, we developed a track map that links material type to the end use for EESDs. The map summarizes the connection between types of material, EES application and 3D printing technique compiled from literature reports cited in previous sections, 3D printer manufacturer datasheets and data analysis[151]. For example, the type of material (insulator, conductor, or semiconductor) can be tracked to optimum printing method, depending on whether it is a functional or structural material in the printed EESD.

All EESD components can be fabricated using any of the common 3D printing techniques in principle. This opens an avenue for additive manufacturing (AM) of fully 3D printed devices. However, not all AM techniques are equally good to produce different EESD parts and materials. For example, due to the chemical and electrochemical properties of Li-ion battery active materials, the cell casing/enclosure must be airtight and solvent-resistant, which requires impermeability and negligible porosity of the printed parts, and chemical stability against softening, polymeric dewetting or volumetric expansion in the case of solvent uptake. Printing techniques based on MJ and ME methods might in some cases be unsuitable for the fabrication of device casing. On the contrary, the 3D printed active material compositions should ideally be highly porous to increase the contact area of the active material and electrolyte and to minimize the entrapment of active particles within the polymer matrix, which makes MJ and ME methods more promising than VAT-P in this case. Thus, the *porosity* of prints produced with different AM techniques is an important factor to consider during design development of 3D printed EESDs.

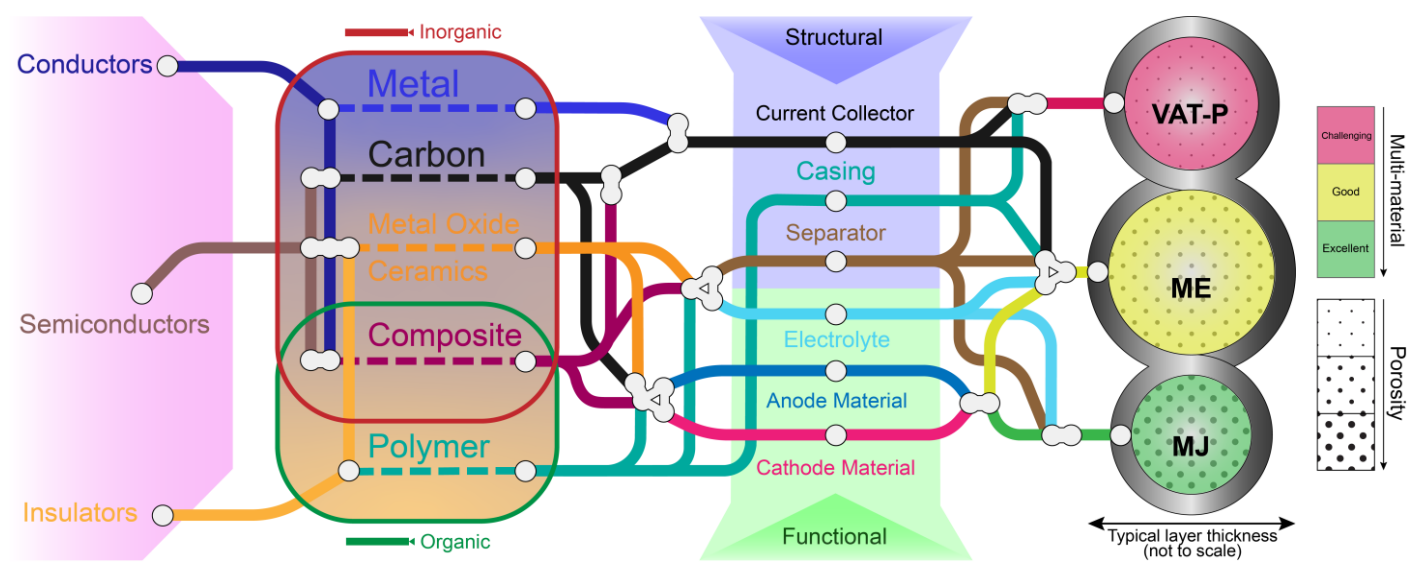

**Figure 4.** Track map demonstrating the connection between different types of materials used in additive manufacturing of EESD, their applications and 3D printing techniques, typically employed for the fabrication of different parts of the devices. The colors of the track lines correspond to the respective materials, material types, applications and printing methods. Junction points indicate multiple materials or composites, applicable to one or several component parts, for one or more 3D printing technique.

*Resolution* in 3D printing technique, which is conventionally defined by distinct parameters such as XY-resolution, Z-resolution (layer thickness) and minimum feature size, plays a notable role in fabrication of high precision, microscopically patterned, porous and hierarchical structures that show great promise in the field of electrochemical energy storage. MJ and VAT-P methods generally perform better in terms of resolution than many other techniques. However, given the continuous improvements in 3D printer hardware, some ME setups (such as FDM printers) could successfully compete with VAT-P methods, being significantly faster at the same time.

*Multi-material 3D printing capability,* i.e. the ability to use different source materials simultaneously or sequentially during the fabrication process, is essential for the creation of fully 3D printed designs and in-plane EESD configurations (see section 3.3). Material extrusion and material jetting methods generally have a good multi-material printing capability, whereas for VAT–P methods (e.g. SLA, DLP) it is challenging, because these methods usually require large amount of photocurable material in a separate tank, and any given layer can be built of only that material. Addition of other materials to subsequent layers would require changing the resin in the tank or tank replacement and, possibly, intermediate post-treatment/cleaning of the partial build. The automation of this process significantly increases the complexity of setup and printing duration but nevertheless can be implemented[152].

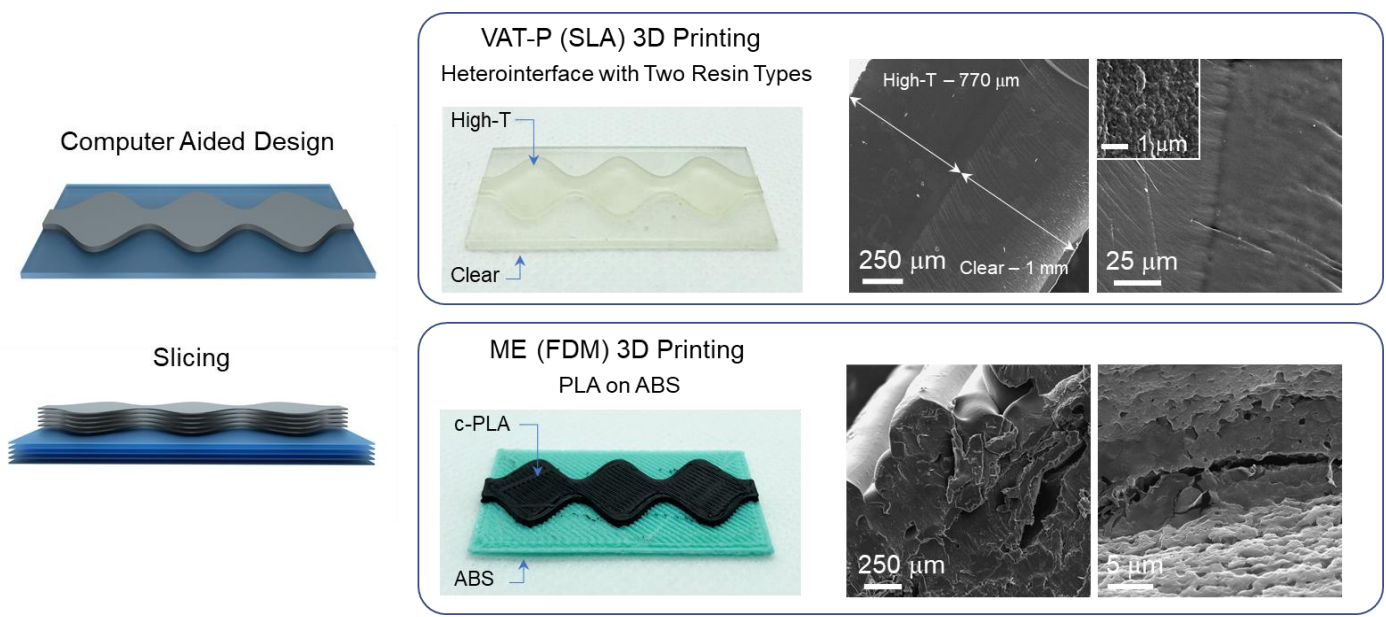

**Figure 5.** Comparison of 3D printing a simple bilayer object using FDM and SLA methods using two commercial 3D printers: MakerBot Replicator 2X (FDM) and Formlabs Form 2 (SLA). The CAD model and its slicing (left) are followed by 3D printing using two different materials (clear and high-temp methacrylate-based resins for SLA; ABS (green) and conductive poly(lactic acid) (black) for FDM). Optical images of the resulting prints are shown (middle). Corresponding SEM images of cross-sectional cuts at different magnifications are presented (right). As modelled, both material layers had the equal thickness of 1 mm. Z-resolutions (layer height) were 100 µm (maximum supported by the printer) and 250 µm (minimum printable) for SLA and FDM, respectively.

In order to compare two widely used 3D printing methods, FDM (a ME process) and SLA (a VAT-P process), we developed test designs and 3D printed identical sandwich-type bi-material objects (see Figure 5). As previously mentioned, the process of 3D printing an object involves several steps. The first step is making a model of the object using CAD or 3D modeling software. This CAD model is then sliced into a stack of planar layers used for conversion of a 3D object to a set of instructions for the printer. After successive printing and post processing the object is ready for application (Figure 5).

3D prints of bi-material hetero-interfaces using two different materials by SLA and FDM show particular characteristics and differences. First, SLA proved much more accurate than FDM in reproducing the modelled structure and dimensions. A significant anisotropy and porosity were observed for the FDM-printed object, which is commonly known, but the improvement in deposit uniformity and quality is consistently better using SLA printing, as shown in Figure 5, the resulting materials and interfaces were more homogeneous and isotropic. FDM printing from thermoplastics such as ABS and PLA, exhibit rough surfaces with considerable porosity on the surface of, and in between, the filaments that comprise the printed object. In addition, the nature of the filament-based printing ensures porosity throughout the macroscale object. By contrast, we find that an identical model printed by SLA, accurately eliminated porosity, and maintains finer definition of the photo-cured resins.

The conductivity (electronic and ionic), surface area (porosity) and mechanical stability are among key parameters that define the performance of 3D printed functional materials in energy storage applications. As a rule, 3D printed functional polymer-based composites exhibit insufficient conductivity, often poor mechanical stability and significant anisotropy (for ME). The current strategy for ME-based printing uses graphite-loaded PLA as a feed material to boost electrical conductivity. Useful efforts have been made to rationally control the conductivity by increasing volume fraction of graphite loading, with a caveat that the material becomes very brittle. The tradeoff is to ensure a mechanically stable printable composite that is more ductile, and this consideration is also valid when loading PLA with active battery materials for example[36,38]. Due to the complete isolation of a significant amount of active material within the insulating polymer matrix, suppressed specific capacity/capacitance and poor C-rate performance are frequently observed. Where the presence of polymer matrix is essential to achieve the material printability and controlled rheology, we provide some considerations for printable electroactive composite formation and discuss ways to improve the performance of 3D printed components target for EES applications. In Figure 6, we address fundamental aspects of PLA-based composites compared to photopolymerizable composites used for 3D printing active materials and/or conductive structures/current collectors/electrodes in batteries or supercapacitors.

In graphite or inorganic powder-loaded PLA (Figure 6), the filler loading is limited (particularly at standard 8.25 wt% loading without using plasticizers). A limited exposed surface area is accessible to electrolyte for supercapacitors, or to active materials if slurry cast over the print. Second, much of the printed structures surface area is electrically passive. A similar situation results when inorganic battery materials are loaded into PLA directly. Electrolyte-material interfacial reactions are limited to those exposed at the top surface of the printed structure, and in the absence of graphitic additive, the printed composite has a very limited electronic conductivity. Ionic diffusivity is also extremely limited in thermoset printed PLA, ensuring that access of Li-ion, for example, to the internal volume fraction of active material is unlikely, severely limiting the gravimetric energy density from inactive mass and reduced voltage (higher resistance). Solvent-induced decomposition is a obvious route to relieving more near-surface graphite and/or active materials in a PLA composite and has been reported in 3D printed PLA-based electrodes in water splitting experiments[153,154]. While expected impurities from a commercial grade PLA undoubtedly lead to some electrochemical activity during the measurements, it is the increase in surface area that leads to some improvements in battery electrodes. On the other hand, increasing the surface area can negatively affect the EESD performance due to the expanded contact between active materials and electrolyte, which has been discussed earlier, and adding porosity to increase interfacial surface area depends on the nature of the electrochemical reaction of choice. It should be noted that conductive additives such as CNTs, may act as fortifiers that counter the onset of brittleness, a strategy used commonly in carbon-fibre composites.

However, care must be taken to ensure PLA-based composites are stable in organic, polar or aprotic solvents so degradation or solvent swelling do no occur. Thermal degradation can also provide similar benefit to solvent etching, with some reports demonstrating that more complex structures are maintained after mild thermal treatment or surface-selective melting. As Figure 6 shows for PLA-based prints, electronic conductivity trades with brittleness, but ionic conductivity always remains limited. For supercapacitor, where conductive surface area is important, some strategies to maximize volumetric porosity form decomposition, or adding complexity to the print design while maintain acceptable mechanical stability, might create more favorable diffusion pathways for electrochemical applications.

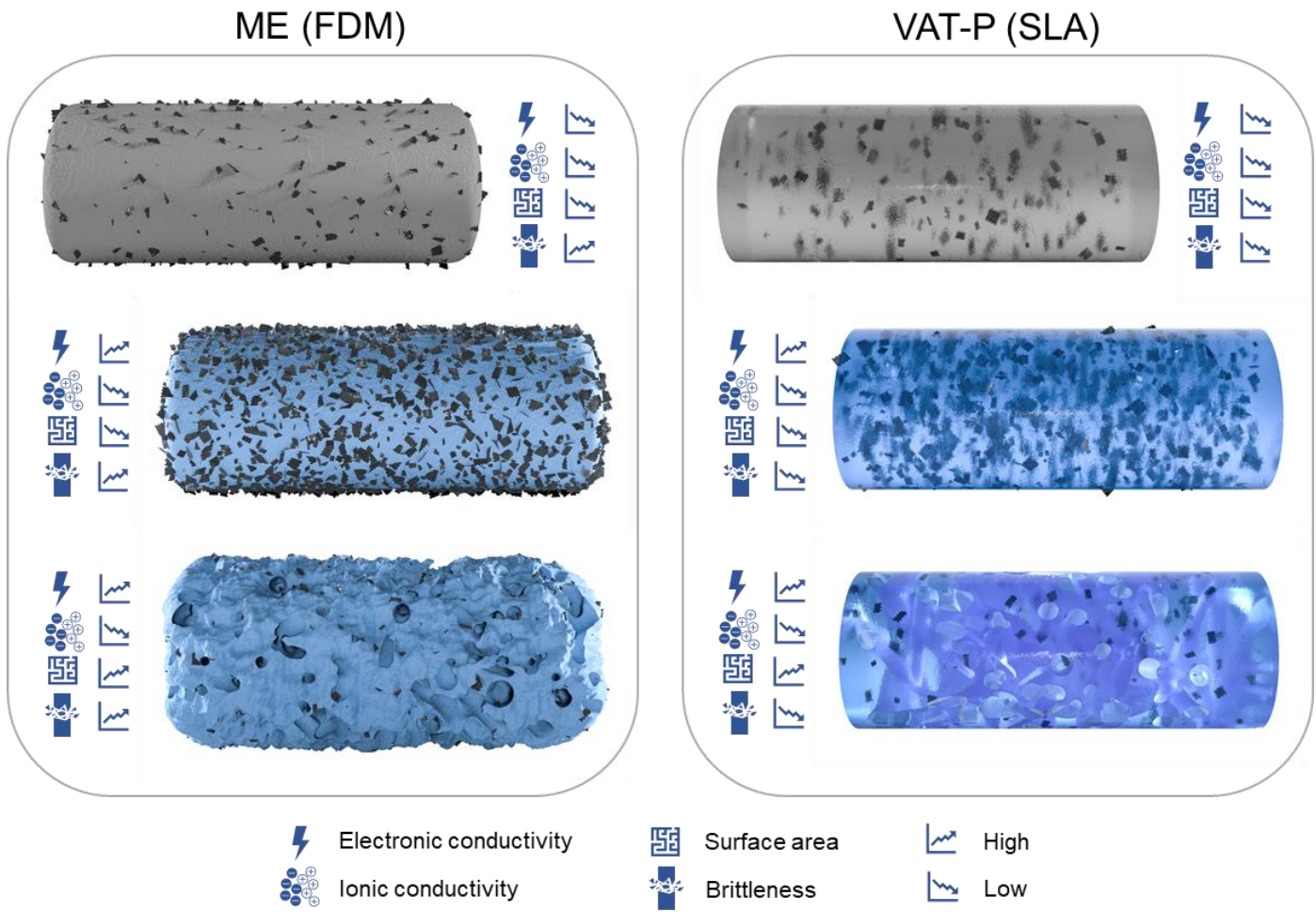

**Figure 6.** Visual illustration of modifications to polymer-based 3D printed composites that affect performance-related physical and electrochemical characteristics. Graphite and/or active inorganic material-loaded PLA (as a pertinent example) thermoset and extruded by an ME printing process can be modified by solvent, thermal or acid/base decomposition. Increasing inorganic volume fraction improves electronic conductivity at the expense of ductility. Inducing porosity increase surface area in a print that is already highly porous (filament structure). Using photopolymerizable resins (VAT-P process), the primary improvement are the mechanical properties, with reduced brittleness, higher resolution, and controllable mass loading being key benefits.

For VAT-P printing methods, of which SLA is a popular system, opportunities exist to significantly improve the nature of active composites or conductive composites. SLA-based methods make possible a much wider range of cured resins at far higher resolution, and the possibility for significantly higher mass-loading with inorganic additives without brittleness (Figure 6). Ionic diffusivity is again limited, but we posit that fluoride-based resin may be developed similar to Li-ion conducting polymer electrodes that may offer a route to truly printable batteries that are mechanically stable, solvent stable, ionically and electrically conducting, and capable of high mass loaded structures.

The comparison of ME (FDM) and VAT-P (SLA) 3D printing techniques presented above demonstrates that the FDM bi-material prints have higher macro- and microscale porosities with an observable gap between the two material layers, and such considerations are valid for single materials prints also. As a consequence, this may lead to the increased interfacial resistance in the case of heterogeneous functional layer stacks, i.e. casing and current collector, or current collector and printed active material structures. By contrast, the bi-material interface in the SLA printed parts is uniform and compact, which may be beneficial for fabrication of multilayer sandwich-type EESDs with very low interfacial resistance. We propose that SLA printing could facilitate the fabrication of active material composites and separation structures with precise, favorable and controlled geometries, good mechanical properties, low internal resistance and high active material content.

## 4. Performance Comparison of 3D Printed Cells

### 4.1 Extrusion, Inkjetting and Stereolithographic Printing of Li-ion Battery Electrodes

In energy storage research, predominantly for batteries and supercapacitors, 3D printing methods and their analogs are gaining some traction. There are several primary objectives when incorporating a new production, fabrication, or composite coating methods to a well-established battery or supercapacitor device/electrode preparation. While 3D printing methods, electrode structure and aspects that influence printing choice for certain materials and applications were summarized earlier, we overview and compare the recent advances made using the three most common printing methods (at the time of writing) for Li-ion batteries and supercapacitor devices. The initial reports using ME, MJ and VAT-P, commonly referred to as fused deposition modeling (FDM), inkjet printing (IJP) and stereolithographic apparatus (SLA), respectively) have been used to fabricate substrates, thin film electrodes and electrolytes in half-cell and full-cell Li-batteries. As shown in Figure 7, the majority of systems reported so far are half-cell Li-batteries using electrodes mainly involving graphene,[155-157] $Li_4Ti_5O_{12}$,[158] $SnO_2$,[159] $MnO_2$,[160] $Si$[161] as anodes, as well as $LiFePO_4$[158,162,163] and $LiCoO_2$ cathodes[164,165]; the handful of full-cell 3D printed Li-electrodes focused on lithium iron phosphate cathode and lithium titanate anode.[34,166,167]

Compared to ME and MJ methods, we are aware of just one report (at the time of writing) using the VAT-P method to print Li-battery electrodes.[34] In this work, Cohen et al. designed a 3D-printed perforated polymer substrate with various shapes and sizes using the so-called SLA technology and further fabricated a tri-layered structure comprising the LFP cathode, $LiAlO_2$-PEO membrane and LTO-based anode by electrophoretic deposition (EPD).[34] This eliminated the need for any metallic materials used as current collector. When cycled from 0.1 to 10 C, a high areal capacity of 400-500 µAh cm$^{-2}$ is obtained for this 3D-LFP cell on perforated graphene-filled polymer substrate. Although the preliminary electrochemical performance of the quasi-solid full 3D mircobattery on 3D polymer substrate suffers from severe capacity decay, the areal energy density of these full-cells was 3× that of the commercial planar thin-film battery. While VAT-P or SLA methods are clearly in their infancy for multistep or single-print electrode fabrication, the authors envisage these kinds of 3D-printed full 3D mircobattery may outperform the state-of-the-art planar thin-film battery if ultra-thin printing of mechanically robust, electrochemically active mass-loaded composites can be printed sequentially.

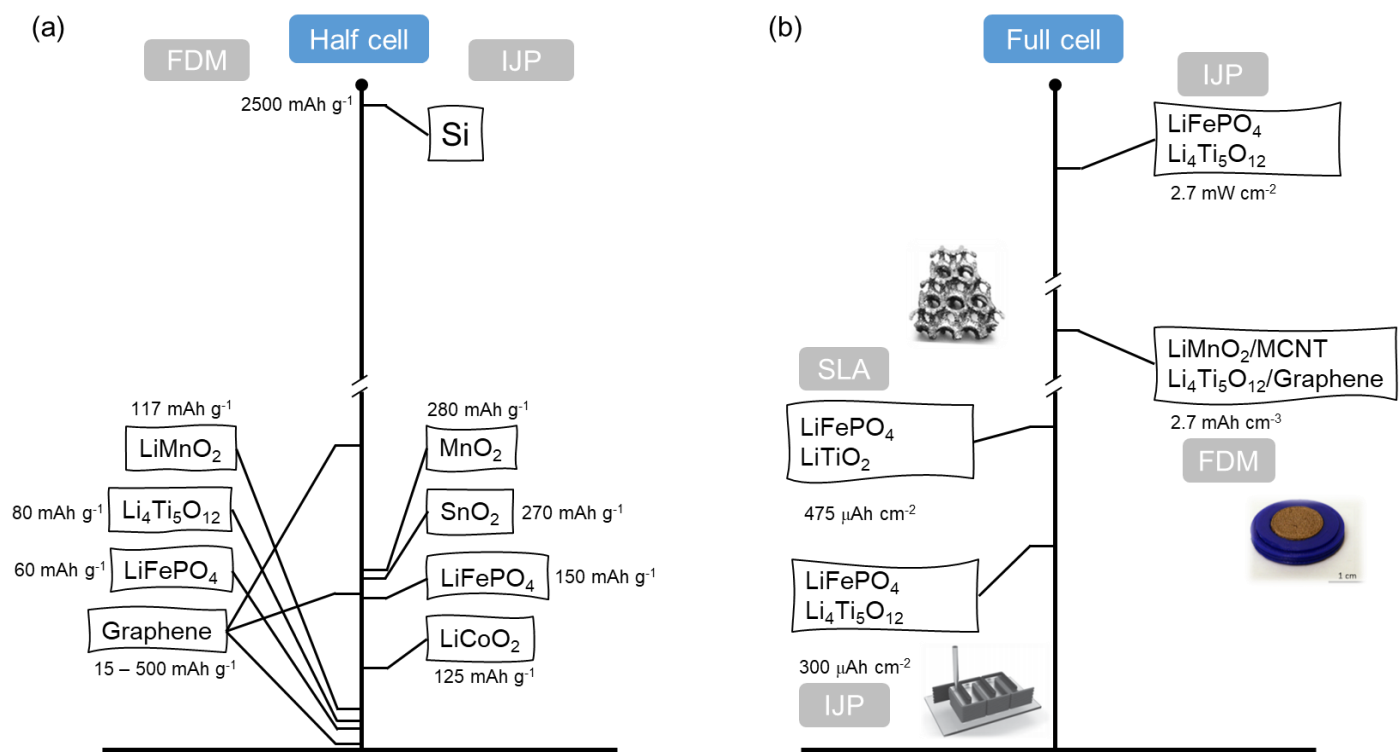

**Figure 7.** Comparison in reported performance metrics for Li-ion half cells and full cells printed using ME (FDM), MJ (IJP) and VAT-P (SLA) 3D printing methods. In (a), the data are shown as reported in mAh g$^{-1}$ and are referenced in the main text. In (b), three unit scales are shown with metrics reports as areal power density, volumetric capacity and areal capacity. These systems are also explicitly referenced in the main text. Representative images from some papers reporting on full Li-ion cells printed using SLA, IJP and FDM are also shown next to the relevant metric.

For MJ printing, of which ink-jet printing (IJP) is the most commonly used, three recent reports highlight advances made in printing full Li-ion cells with LiFePO$_4$ (LFP) cathode and Li$_4$Ti$_5$O$_{12}$ (LTO) anode, and these cells exhibited excellent lithium storage performance. Representative work is that the IJP printed 3D microbatteries composed of LTO and LFP microelectrode arrays in an interdigitated architecture carefully designed by Sun et al. offer high areal energy density of 9.7 J cm$^{-2}$ at a power density of 2.7 mW cm$^{-2}$.[166] In addition, Delannoy et al. demonstrated the IJP silica-based ionogel onto iron phosphate cathodes and titanate anodes as porous composite electrodes, respectively, which were assembled together with a solid state 3D IJP printed ionogel electrolyte, and the full Li-ion cell showed a areal capacity of 300 mAh cm$^{-2}$ for up to 100 cycles.[167] This 3D IJP printed silica based ionogels are shown processable for microbatteries as improved safe, cost effective, high ionic conductivity and thermal resistant electrolytes. Furthermore, considering the evaporation and possible leakage of conventional electrolyte to cause safety concern and capacity decay in battery, Fu et al. used a polymer composite ink containing a mixture of poly(vinylidenefluoride)-co-hexafluoropropylene (PVDF-co-HFP) and $Al_2O_3$ nanoparticles to formulate a solid-state composite, which served as the electrically insulating separator as well as the gel polymer electrolyte. At the same time, using water as a greener solvent made an aqueous GO-based electrode composite ink system to obtain LFP/GO cathode and LTO/GO anode. When assembling an arranged interdigitated pattern, the ink composite consisting of PVDF-co-HFP and $Al_2O_3$ nanoparticles was printed into the channels between two electrodes. This full cell showed a capacity of about 100 mAh $g^{-1}$ after ten cycles at a specific current of 50 mA $g^{-1}$, and the coulombic efficiency increased to nearly 100% from the 2$^{nd}$ cycle.[32] These demonstrate the feasibility of 3D MJ full-cell Li-ion batteries with competitive performance compared to conventional Li-ion batteries. We note that MJ processes are long-established and its integration into additive manufacturing has grown considerable in recent years, but its layer-by-layer methodology does not allow for the versatility of CAD-based form factor design possible with other 3D printing methods. 3D FDM printing was attempted for full lithium ion batteries by Reyes and co-workers that they fabricated the LTO/Graphene anode and LMO/MCNT cathode by mixing Poly(lactic acid) (PLA) polymer, conductive addictive and active material. It demonstrates the use of these novel materials in a fully 3D printed coin cell, which can exhibit an average volumetric discharge capacity of 3.91 mAh $cm^{-3}$. Benefiting from 3D printing with the ability to print arbitrary shapes and sizes, moreover, it is very interesting and meaningful that the FDM printing integrated printed batteries eventually were used in the production of 3D printed wearable electronics including the LCD sunglasses, the LCD panel as well as LED bangle.[38]

For Li-ion half cells, FDM printing technology was successfully employed to prepare graphene, $Li_4Ti_5O_{12}$ and $LiFePO_4$ electrodes so far. Foster et al. were the first to report lithium storage performance of ME printed 3D disc electrode architectures (made of 8% graphene and 92% poly(lactic acid, PLA).[155] However, because of the low electrical conductivity of PLA disc electrode, the discharge specific capacity was limited to 15.8 mAh $g^{-1}$ at a specific current of 40 mA $g^{-1}$. The next iteration to this ME-based 3D printed PLA current collector approach, was to relieve the internal graphite component of the conductive PLA to increase the effective electrochemically active surface area. As detailed in Section 3.4, conductive plastics suffer from low electronic and ionic conductivity. While graphene or graphite addition to PLA above a percolation threshold as a mass fraction, can improve electrical conductivity, Li ion diffusivity through the PLA to all graphite is severely limited. Foster et al. recently fabricated FDM printable graphene/PLA filaments with higher graphene content, and the specific capacity was significant improved after chemical pre-treatment (500 mAh g$^{-1}$ at current density of 40 mA g$^{-1}$), which induced surface porosity to increase the available surface graphite, with some improvement in performance of the half cell anode.[156] Comparatively speaking, 3D electrodes prepared by MJ printing methods shown better half cell performance compared to PLA, ABS or plastic counterparts. This is directly a function of the nature of active material composite, where in MJ processes that active material and/or conductive additives can be printed directly, rather than as a composite with plastic support materials as is common in ME printing. For example, MJ prints using LiFePO$_4$ cathodes can deliver a high specific capacity of 151 mAh g$^{-1}$ at current density of 15 mA g$^{-1}$.[162] Even carbon-coated LiFePO$_4$ cathodes printed via MJ technology exhibit 80 mAh g$^{-1}$ at rate of 1530 mA g$^{-1}$ without significant capacity decrease for 100 cycles.[163] In addition, for the inkjet printing of LiCoO$_2$ thin films, an initial discharge capacity of 120 mAh g$^{-1}$ is reduced by only 5% after 100 charge–discharge cycles, at a current density as high as 384 μA cm$^{-2}$.[164]

These reported findings summarise the pros and cons of 3D printing for electrodes and cells. By MJ processes, one can design shape, form factor and areal loading of functional active material composites directly, without the complications of low ionic conductivity host plastics as the build material common for ME methods. By contrast, ME printing allow the design and printing of entire objects, including current collectors, electrodes and casing in sequential or single step printing format, which is challenging for MJ or inkjet processes, even for current-collector-free approaches. High performance full-cell and half-cell Li ion batteries using 3D printed electrodes will require optimization of the composite mixtures to improve ambipolar conductivity of a current collector and electrode material formulation, rheology of deposition. Another practical challenge for Li-ion cells is the stability of many plastics with organic electrolytes and solvents, which is somewhat less problematic for 3D printing of supercapacitors, which we discuss next.

**4.2 3D Printed Supercapacitors**

Complementary to Li-ion batteries, 3D printing approaches have also been used to develop supercapacitor-type energy storage devices (Figure 8). In principle, these devices are easier to fabricate, are less susceptible to ionic conductivity limitations of the internal bulk volume of building materials, and various polymeric or plastic materials are more stable in aqueous-based electrolytes. Issues related to limited electrical conductivity in build materials remain, culminating in rather large internal ohmic losses. Rational control of porosity or high surface area required for many high capacitance supercapacitor systems using electrical double layer charge storage processes, and thus far, degradation of the building material surface in solvents, alkaline or acidic solutions has been used to increase the geometrical surface area. These approaches, and identification of contaminants that are expected in low cost PLA, have been more common in plastic-based prints using ME processes and tested for their effect on water oxidation or hydrogen evolution electrochemistry.[168-170]

A variety of electrode materials have been explored for supercapacitors including carbonaceous materials, 2D MXene and metal oxides (Figures 2 and 8). MJ (i.e. inkjet printing) is one of the most popular technologies for electrode preparation because complex patterns and electrode geometries can be programmed for jetting or printing. Additionally, MJ provides a higher resolution and multi-material printing capability that are more difficult to achieve by self-assembly or directed-assembly approaches of materials or composites onto patterned electrode substrates. So far, most reports using MJ methods have been symmetric supercapacitors. One interesting development was the flexible solid-state asymmetric supercapacitor system using MJ technology, based on lamellar $K_2Co_3(P_2O_7)_2 \cdot 2H_2O$ and graphene nanosheets. This device delivered a relatively high volumetric capacitance of 6 F $cm^{-3}$ and had excellent cycling stability (5.6% capacitance loss) after 5000 cycles at 10 mA $cm^{-3}$. Owing to the layering capability of MJ processes using inks with a high volume fraction of active materials, this device achieved a maximum volumetric energy density of 0.96 mWh $cm^{-3}$ and power density of 54.5 mW $cm^{-3}$ at a rate of 100 mA $cm^{-3}$, which is superior to most solid-state micro-supercapacitors.[171] MJ printing of asymmetric supercapacitors compositing of $MnO_2$/Ag/MWNT anode and MWNT cathodes also competes with other $MnO_2$/metal/graphite-type supercapacitor configurations that are known to have very long cycle stability. This device exhibited a capacity retention ratio of 97% over 3000 cycles, and high volumetric energy density (1.28 mWh $cm^{-3}$) and power density (96 mW $cm^{-3}$).[172] These recent reports highlight several features that are possible by MJ, particularly the ability to have high mass loading of multimaterial systems, to print in thin film format in more complex geometries and enable good registry between cathode, electrolyte and anode prints, removing mismatch in mass, printed area and thickness differences in thin microcapacitor devices. MJ still cannot enable a full complete cell for either Li-ion or supercapacitor technology, and casing in a single print format remains a challenge compared to ME or VAT-P processes.

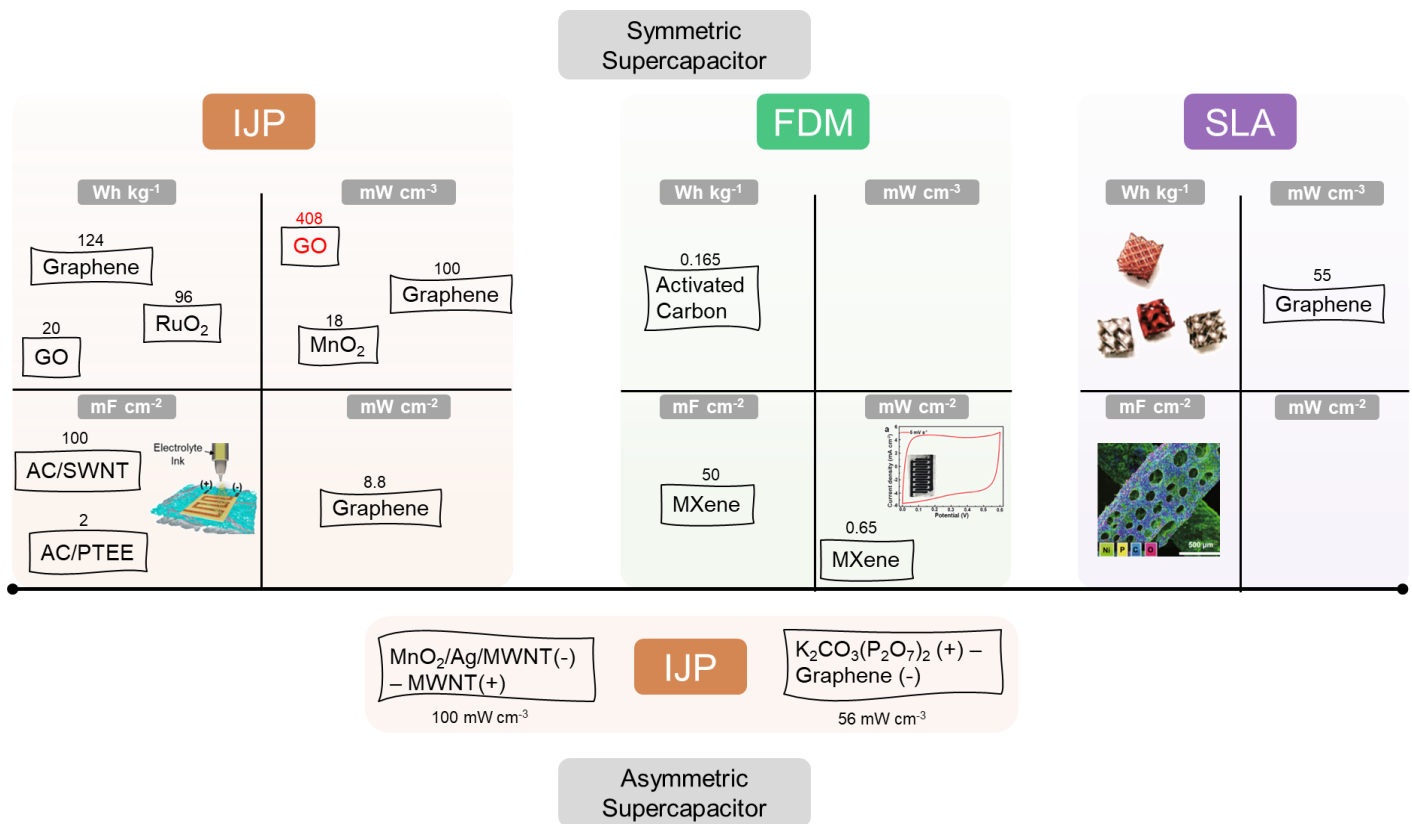

**Figure 8.** Map of recent developments and performance metrics for symmetric and asymmetric supercapacitors prepared using a range of materials by MJ (IJP), ME (FDM) and VAT-P (SLA) 3D printing processes. The reported values in gravimetric energy density, volumetric power density, areal capacitance and areal power density are provided in each color-coded region for symmetric supercapacitor systems. Further details to referenced publications are described in the main text.

Compared to asymmetric supercapacitor applications, MJ-based electrodes for symmetric supercapacitors have unsurprisingly received more attention and made more progress. For example, GO-based,[173,174] graphene-based[175-177] and activated carbon (AC)-based materials[178,179] via MJ methods have reported to apply in symmetric supercapacitors. Le et al. found that hydrophilic GO nanosheets dispersed in water form stable inks that could be inkjet printed on Ti metal current collectors, delivering a specific capacitance of between 48 to 132 F g$^{-1}$ using scan rates from 0.5 to 0.01 V s$^{-1}$.[173] Inks made from GO are commonplace, and thermal or photothermal reduction can be used to render printed graphene electrodes. In another example, introducing porosity into MJ printed GO electrode by photothermal reduction, interdigitated GO on flexible substrate was designed, achieving a volumetric power density of 0.408 W cm$^{-3}$ in an ionic liquid electrolyte. These performance metrics are comparable with commercial supercapacitors.[180] Chi et al. developed an all-solid-state symmetric supercapacitor based on inkjet printing a graphene hydrogel-loaded polyanilic (GH-PANI) electrode, to give a power density of 0.4 kW kg$^{-1}$ and energy density of 24.02 Wh kg$^{-1}$.[181] Li et al. demonstrated that that printing graphene/DMF dispersions as inks on the fingers of interdigitated structure in a symmetric supercapacitor deliver a reasonably high areal power density of 8.8 mW cm$^{-2}$.[177]

Taken together, these investigations using graphene-based dispersions and polymeric inks offer useful ways to uniformly thin supercapacitor electrodes. As a layer-by-layer method, MJ offers exquisite control over thickness and when exfoliated graphene inks are printed this way, a power density of 124 Wh kg$^{-1}$ with an associated energy density of 2.4 Wh kg$^{-1}$ has been shown for a symmetric supercapacitor.[176]

In order to print metal oxide electrodes from particulate ink dispersions, SWNTs or MWNT were added to inks to improve the electrochemical performance by increasing the internal conductivity of the resulting printed material. For example, a $RuO_2$ nanowire/SWNT hybrid film electrode for a symmetric supercapacitor exhibited a power density as high as 96 kW kg$^{-1}$ by adjusting the rheology of the SWNT and $RuO_2$ NW ink mixture to print a conformal and conductive capacitive electrode.[182] The same approach was also used by Lee and co-workers who fabricated an SWNT/AC electrode by direct inkjet printing on conventional an A4 paper sheet. The cyclic voltammetry (CV) profiles appeared to be nearly rectangular in shape, and an areal capacitance of 100 mF cm$^{-2}$ over 10,000 cycles without any significant capacitance loss was reported.[179]

By comparison, there are fewer publications using ME based methods (such as FDM) to print electrodes for supercapacitors. In early work, 3D printed electrodes with quite a low mass of graphene were studied in a solid-state supercapacitor, exhibiting a low capacitance of 28 µF at 0.5 µA.[155] ME-printed AC/flexible fabric electrodes for supercapacitors have been attempted, achieving an energy density of 0.019 Wh kg$^{-1}$ and a power density of 165 W kg$^{-1}$.[183] Yao et al. adopted ME technology to print 3D substrates for PPy/rGo nanocomposite deposits, and further developed a symmetric solid-state supercapacitor that delivered 98.37 F g$^{-1}$.[184] Given the ubiquity of low cost ME-type 3D printers available to laboratories and the general public, it is somewhat surprising that reports on supercapacitor configurations are rare, even those involving relatively simple chemistries of alkaline aqueous electrolytes and high surface area carbons in symmetric supercapacitor form factors. Just recently, some strides are being made with MJ approaches involving plastic building materials, namely the use of concentrated inks to increase thickness and mechanical robustness in so called current-collector-free systems. By printing three-dimensional MXene architectures in three dimensions as the building materials (as opposed to typically used ABS or PLA), the electrodes can show excellent supercapacitor performance. Zhang et al. demonstrated that ME printing of interdigitated MXene-based electrodes for symmetric solid-state supercapacitor showed promising performance with an areal capacitance of 61 mF cm$^{-2}$, excellent rate capability with 82% retention (5-200 mV s$^{-1}$), and a long lifespan of more than 10,000 cycles, as well as superior energy density of 0.76 mWh cm$^{-2}$ and power density of 0.63 mWh cm$^{-2}$.[185] In a similar approach, additive-free 2D $Ti_3C_2T_x$ inks were layer by layer printed to form freestanding, electrodes with high specific surface area architectures of different sizes and shapes. [186] This current collector-free approach uses the active material as the building material in an ME process directly, addressing the challenge of ionic diffusivity issue to some degree, while improving the electrical conductivity since MXenes are highly electronically conductive. When they were assembled into a symmetric supercapacitor, a quasi-rectangular CV curve shape was presented, showing an ideal capacitive behavior. The areal capacitance for this approach can attain 2.1 F cm$^{-2}$ at 1.7 mA cm$^{-2}$ with capacitance retention of 90% after 10,000 cycles. This translates to a significant energy density of 0.0244 mWh cm$^{-2}$ and a power density of 0.64 mW cm$^{-2}$.

The recent advent of VAT-P methods, or SLA printing as it is more commonly known, has moved from industrial prototyping to laboratories and to hobbyists as the cost of SLA printers and associated resins has become more accessible. We envisage VAT-P methods being used as the next step in energy storage device and materials research. Additive manufacturing here involves multi-material direct printing with sub-mm or tens of μm level resolution with a wide range of photopolymerizable resins. At the time of writing, we are aware of one report using VAT-P printing to produce hierarchical graphene in quasi-solid symmetric supercapacitor devices. By photopolymerizing a hierarchically porous graphene a high areal capacitance of 57.75 mF cm$^{-2}$, good rate capability (capacitance retention of 70% from rates in the range 2-40 mA cm$^{-2}$), coupled with long cycle life (capacitance retention of 96% after 5000 cycles) have been achieved. The maximum power density maintained at 12.56 mW cm$^{-2}$ (56.52 mW cm$^{-3}$) with a power density of 0.0061 mWh cm$^{-2}$ (0.027 mWh cm$^{-3}$), stated to be comparable to the state-of-the-art carbon-based supercapacitor.[41] Because of inherent disadvantage of the conventional SLA method with long fabrication duration and large beam size, digital light processing (DLP) is introduced as faster VAT-P printing technique to design supercapacitor electrode. Recently, DLP printing-based hierarchically cellular lattices were built with different structures. These metallic lattices may serve as the current collectors or conductive scaffolds because of superior conductivity of 2 Ω cm$^{-2}$. 3D hierarchically porous graphene on octet-truss metallic lattice was demonstrated, which was beneficial to electrolyte penetration and ion diffusion. When applied in the symmetric supercapacitor device, the areal capacitance of 57.75 mF cm$^{-2}$, long lifespan of 96% after 5000 cycles as well as superior energy density of 0.008 mWh cm$^{-2}$ were obtained.[41] Similarly, DLP printed well-defined 3D hierarchical micro-supercapacitor electrode composed of durable octet micro-trusses exhibited high surface area (2931 mm$^2$ g$^{-1}$), high conductivity, and a specific capacitance of 3.01 mF g$^{-1}$.[37]

In addition, using other printing methods such as laminated object manufacturing (LOM)[33,187], binder jetting (BJ)[188], powder bed fusion (PBF) or selective laser melting (SLM)[30,189,190] for fabricating 3D electrode in energy applications have been sporadically reported. These approaches producing 3D frameworks or hierarchical nanostructures lead to the maximization of electrochemical performance in batteries or supercapacitors. For example, selective laser sintering method was used to fabricate metal scaffolds with controllable porosity by adjusting laser power and scan speed in the work by Liu et al. The different structures can be optimized for better charge carrier mobility and increased electroactive surface area, which result in lower ohmic resistance, faster charge transfer and mass transport. Furthermore, the optimization of the 3D printed structure improves cycling lifetime and capacity of the pseudocapacitive material.[189]

As three most potential 3D printing methods, IJP, FDM and SLA technologies present unique capabilities in the electrochemical energy storage systems including Li-ion battery and supercapacitor. In particular, the design of various shapes and structures can produce controlled porosity and abundant exposed active surfaces, which are beneficial to fast ion diffusion and electrolyte penetration and particularly to maximize the performance per unit mass or volume. However, these technologies currently sufferd from limitation towards the practical application. Standard 2D IJP methods prove difficult to obtain multi component structures, high resolution, and large volume prints, and so is less suitable for commercial devices. Although FDM or SLA technology is capable of building up large system, the limited accuracy in FDM and the challenges in multi-material printing by SLA currently inhibit straightforward applicability in energy storage sectors. The other printing methods are introduced to fabricate 3D electrode because of these manufacturing limitations. Future efforts are required to develop 3D printers able to combine multi-features of different printing method or develop new multi-function 3D printing systems.

## 5. Outlook and future directions

In this review, we offered a perspective on the choice and use of materials, especially feed materials, for 3D printing of any electrochemical energy storage device. The choice of printing method, the requirement for active materials or electrode design, printing an electrolyte with state-of-the-art chemistry, and overall cell design, are underpinned by what is printed/printable, and how it is printed. There are several reviews available on the various reports published in recent years[6,16] where individual performance can be assessed and compared, including the often very disparate methods and materials of printing, but with similar overall goals. Having surveyed this literature in detail, our strategy here was to provide the reader with a perspective and guide to materials choice, electrode design, and printing method in the context of reported values for many printed EESDs, so a rational decision could be made at the outset for a particular device for a dedicated application.

The basis for reconsidering the form factor of energy storage devices is a consideration that puts the use of the device and user at the forefront of the technology. To realise this goal, knowledge of composite structures and battery assembly methods specific to AM and 3D printing that work cooperatively is urgently needed, particularly for miniaturized integrated power sources and energy storage devices for sensors, integrated electronics, wearable and mobile technologies. We acknowledge that many small form factor technologies will require batteries where 3D printing and AM will become very useful, while some important advances in wearable technologies for virtual or augmented reality[191], electronic skin[192], sweat sensors[193] or wireless pharmacology[194], do not require power sources in certain designs. Without a strategy to control the trade-offs between all physical and electrochemical properties of the feed materials and resulting active materials or structures, 3D printing of energy storage devices will not progress. All strategies to improve surface area from the printing method itself (such as Hilbert curve, fractal geometry or other arrangements of the current collectors for example)[195] will need an optimized material to be printed that is entirely electrochemically addressable with well understood and stable interfacial reactions between anode, cathode and electrolyte; this is especially true for 3D printed batteries. Summarised in Figure 9(a) are examples of metallized (copper) SLA-printed structured microlattice (pore-and-truss) current collectors of various interconnected crystalline frameworks to improve intra-electrode conductivity, active material loading and damage tolerance. Recent advancements by Chen et al. show that electrochemical 3D printing[196] can enable metal infused electrospun carbon fibre electrodes to add either pseudocapacitive materials or current collector material into porous electrodes. Such mesh based current collectors mimic metal foams or ultralight metallic microlattices[197], but can be designed to have a high or low density of current collector wiring, or indeed be designed structurally to accommodate significant bending, compression of shape changes. For example, hierarchical structuring, complex geometries or hardening that is inspired by crystal structures (grain boundaries or crystalline periodicity for example)[198], allow for damage-tolerant structures to be realised and printed, which could be useful for some electrode design for EESDs. Tolerance to volumetric swelling in Li-ion materials, negative Poisson ratio structures[199], or engineering porosity and printed trusses within an structure can significantly alter the mechanical properties[200] in addition to maximizing active material loading, or tolerance to physical abuse during usage within a device containing 3D printed components. We point the reader to an informative review on architected cellular materials[201] or indeed biosinpired structural materials[197], where inspiration may be drawn to advance EEDS electrode designs. In principle, this approach can angle ultra-thick electrodes for laminar or in-plane 3D printed electrode design in EESDs.

In our opinion, customizing the shape of the battery or supercapacitor cell should be a goal of AM fabrication of EESDs tailored to the eventual use, and leveraging the power of CAD at the product design level should enable function using printed batteries that has not been accessible before due to a cell's fixed form factor and bulk design. Weight saving, compared to stainless steel, aluminum or copper etc. is not sufficient to transition to plastics; aluminium, titanium or other light metals can provide enough weight saving in principle at full cell level in an alternative form factor design. Mirroring the control and knowledge at the active material and/or construction material level, at least as good as Li-ion systems, will be needed to make a big step in realizing battery or capacitor power match to design and function of system that have not been envisaged yet because of the fixed form factor of existing cell designs.

**5.1 Reproducibility in 3D printed electrochemical energy storage devices**

As the field self-organizes into the future, the development of 3D printing for batteries, supercapacitors and other energy storage devices will require testing of composite or printed material stability in whichever functionally most useful form is best for a particular application. In batteries and capacitor-devices that use organic electrolytes and/or solvents, feed material choice will need good long term stability, to avoid in situ decomposition (leading to mechanical failure, or spikes in electrochemical performance) or deterioration in shape (particularly if fine featured, microlatticed or complex in design) when solvent uptake or dewetting occurs. These effects become more pronounced in prints that are treated to add porosity to control or improve the access to higher volume fractions of active material. We suggest too that the initial strides in 3D printed batteries offer good insight on the need for stability studies for the reasons above, and to ensure reproducible and comparable results for half-cell and full-cells tests. As the primary aim of a 3D printing is to form new shapes of EESD, ideally in a single step, or an uninterruptable sequence of steps, we propose that all cells should be tested in a 'full cell' configuration, i.e. as a printed device. It is possible to add printed electrode materials to half cells or even non-printed full cells to ensure they function correctly, and are benchmarked with a stable system in a particular laboratory. Nevertheless, testing complete cells where casing and current collectors are printable is a necessary step in our opinion. Feed materials for 3D printing influence electrolyte choice, cell sealing considerations, current collector design, and active material composition printability. Various forms of dual active material-containing electrodes, possibly current collector-free, printed together with a new form factor cell casing will require half-cell and full cell testing. In the battery and supercapacitor fields, we do have standard cell testing systems (coin cells, pouch cells, flooded cells or others) that allow reasonably good comparison between materials, systems and research group results. For ME, MJ or VAT-P processes, it may be necessary to propose a freely accessible standard cell design and volume, that provide a common basis for cell-to-cell comparison for the field.

## 5.2 Mixed method approaches for 3D printing

Some of the latest developments in AM and 3D printing may provide a breakthrough in the fabrication of EESD. Since any given battery or supercapacitor requires several materials for casing, separator, electrolyte, active material and current collectors, separate material processing tend to provide better performing systems than composite approaches that attempt full cell fabrication in a single printing step. This might change when new materials are capable of being printed, or multi-material or multi-method system are developed. Figure 9(b) demonstrates a recent 3D printing advance, where polymerization and pyrolysis process were used to convert a miscible photopolymer into a transparent printed glass[202]. With the advent of glassy materials or ceramics for solid state battery electrolytes, further work along these lines may provide options for dense or porous structured and printable glasses and ionically conducting glassy ceramics.

To address the multi-material challenge with voxel level resolution, Skylar-Scott et al. developed a multi-material multi-nozzle 3D (MM3D) printer in which the composition, function and structure of the materials are programmed at the voxel scale.[5] This advances extrusion based technology to allow programmable addition and switching between eight different materials. In the 3D printed battery field, this approach could tackle the requirements for a base structural material with seamlessly integrated conductive material, if the latter can be made as a printable viscous ink suitable for MM3D. While the system is limited to printing the same object in which material types can be integrated by programming the nozzle control, massive parallelization could be possible for high volume production of complex electrode structures and formulations.

Regarding new feed material solutions for EESD additive manufacturing, Cheng et al. recently reported a way to print solid state electrolytes at higher temperatures, that were shown to be compatible in EESDs[203]. The ionic conductivity was on the order ~$10^{-3}$ S cm$^{-1}$, making it suitable as a solid state electrolyte. The approach mitigates solvent evaporation issues that occur when electrolyte and electrode formulations are printed together. Using a nanoscale ceramic filler in a PVDF-*co*-HFP polymer, a continuous dense film could be printed that showed good wetting with active materials to lower the interfacial resistance. The hybrid solid-state electrolyte consists of the solid polymer matrix and ionic-liquid electrolyte. The solid polymer matrix enables Li-ion diffusion that provides sufficient mechanical support to separate both electrodes. Their direct ink writing approach avoided any post-treatment for the printed materials.

Post-treatment procedures (heating or freeze-drying) can cause distortion in 3D printed structures when solvent incorporation was necessary to create the feed material. Shrinkage from solvent evaporation and knock-on effects on material-material interfaces (reduced thermal, electrical and ionic conductivity and mechanical integrity) remain challenges for inorganic electrolytes[204] and also for printable ones, but the high temperature direct writing method bodes well for Li-polymer and solid state microbattery development.

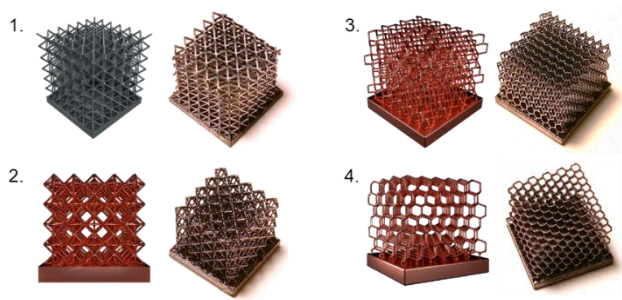
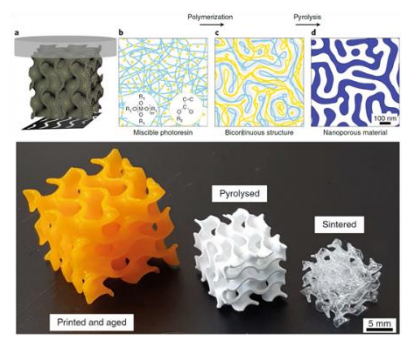
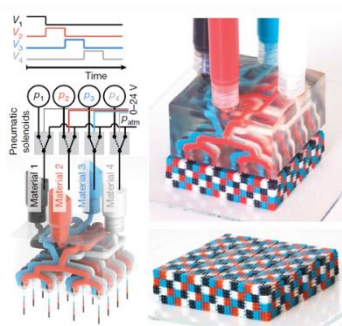
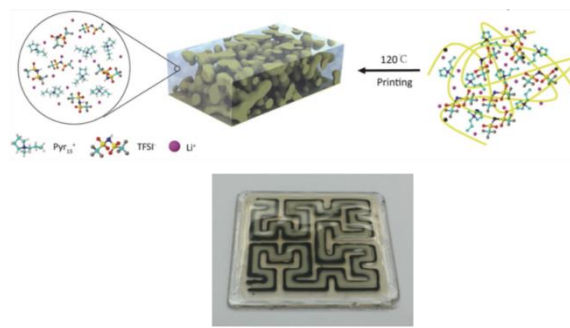

**Figure 9.** (a) Schematic representations and actual VAT-P printed copper metallized microlattice current collectors with four different interconnected structures: 1. open cubic, 2. octahedral, 3. diamond, 4. diamond (rotated). Each structure represents an approach to internal wiring of thick active material electrodes in 3D printed cell electrodes. (b) 3D printing using digital light processing methods (a VAT-P type process) using photopolymerization-induced phase separation of hybrid resins to form complex printed glass with high spatial resolution and multi-oxide chemical compositions. (c) A direct ink writing approach that deals with multi-material requirements in a printed structure. Voxelated soft matter demonstration using multi-material multi-nozzle 3D (MM3D) printing. The method allows composition, function and structure of the materials to be programmed at the voxel scale to give seamless, high-frequency switching between up to eight different materials to create voxels with a volume approaching that of the nozzle diameter cubed. (d) Direct fabrication of electrolyte from printable inks at an elevated temperature using solid poly(vinylidene fluoride-hexafluoropropylene) matrices and a Li$^+$-conducting ionic-liquid electrolyte, which was modified by the addition of ceramic fillers to give an ionic conductivity of $0.78 \times 10^{-3}$ S cm$^{-1}$. Reproduced with permission from Nature Publishing Group (2019), AAAS (2019), and Wiley-VCH (2018).

Finally, similar to multi-method production lines in battery assembly currently, multi-method AM would be one approach to dealing with the trade-offs associated with certain printing methods for EESDs, even those that enable multi-material printing. As it stands, multi-material printing with voxel-level resolution is limited to fixed designs, and certain (usually viscous) materials. Metal 3D printing, glassy ceramic printing, solid material and active material printing by ME, MJ, VAT-P and other printing techniques in a single system would add production line capability to alternative form factor full cell additive manufacturing.

**5.3 3D printed cell viability and safety**

Safety concerns will always be important, and there are formal UN38.3 regulations (among many other regulations from relevant bodies) pertaining to cell design and requirement for transport, and as such 3D printed technologies will require certification to be safe at altitude, under vibration, and under thermal shock. At cell chemistry level, the community is aware of issues related to electrochemical systems such as batteries, where flammable electrolytes and high energy materials are commonplace. The nature of their behavior as a function of charge rate, state of charge, depth of discharge, cycle life etc. are becoming very well understood even with newer higher voltage (>4.5 V) systems. Nominally, materials, composites, electrolytes and their interface to metallic current collectors define a system that is similar in all cases, in terms of cell design and form factor. When the form factor is modified, new testing followed by validation and certification is necessary. This may prove to be the bottleneck for commercial application, even with a promising material set that enables truly bespoke cell designs from ultrathin and flexible, to complex shapes that are seamlessly integrated and essentially invisible with the product design. The flexibility that 3D printing offers juxtaposes the stringent requirements of fixed form factor cells necessary for IEC 61960 testing and certification, as a pertinent example for lithium single cells for portable applications. Cycle stability, self-discharging tests and other safety-performance tests will likely need to be defined and devised for each and every new form factor battery cell, which poses an obvious time and cost burden for customizable form factor EESDs.

It is unclear yet, if new revision to the IEC protocol on accelerated testing of lithium cells, will be applicable to any form of 3D printed cell based on lithium chemistry. Form factor modifications automatically require new certification as existing certification is designed to ensure comparable analysis of cells by different manufacturers. A standard for 3D printing in the battery field, will be useful to compare performance, but the freedom to invent and develop a vast array of mixtures, composites and designs is an issue that will need to be addressed for viable printable battery technologies for the marketplace.

## 5.4 Cycle life, energy density and application

For EESDs designed to maximize volumetric energy density, 3D printed solutions may indeed prove useful. The premise for 3D printing any battery or supercapacitor is predicated on its ability to provide something that existing relatively small volume or ultrathin Li-ion cells cannot provide. Maximizing volumetric energy density first requires a reduction in non-active mass. This is possible with lighter weight, low MW polymers as feed materials, with an optimized volume fraction of any necessary conductive or active materials. The outer casing should also be lightened compared to a miniature pouch cell or stainless-steel coin cell equivalent. With appropriate polymers and materials, stability in high voltage solvents would then provide the optimized energy density. We posit that the development of 3D printable Li-ion cells should be rechargeable, and not single use.

With 3D printing via CAD and high-resolution prints with photoresins (VAT-P) as a working example, the choice of feed materials, and development of complex composites can be decided at the design stage for the product that it would power. For example, an internet of things or 5G array for smart homes, wearables, sensors, etc. will require small cells with long cycle life and stability since the power draw is limited. If such cells are printed to integrate into the outer cover of a smart sensor, matching product specification (colour, form factor, location etc.), then integrated charging with battery management systems becomes one of the challenges. Printable battery or supercapacitor cells can in principle be matched with piezotronics, photovoltaics, or thermoelectrics to continuously charge the cell under low power demand requirements.

**Acknowledgements**

We acknowledge support from the Irish Research Council Advanced Laureate Award under grant no. IRCLA/2019/118. This work was financially supported through SmartVista project, which has received funding from the European Union's Horizon 2020 research and innovation programme under the grant agreement No. 825114. Support from Science Foundation Ireland under grants no. 15/TIDA/2893, 17/TIDA/4996 and 14/IA/2581 are also acknowledged. Additional support is also acknowledged from